\documentclass{amsart}

\usepackage[margin=1in]{geometry}
\usepackage[foot]{amsaddr}
\usepackage{algorithm}
\usepackage{algpseudocode}
\usepackage{bm}
\usepackage{booktabs}
\usepackage{enumitem}
\usepackage{graphicx}
\usepackage{hyperref}
\usepackage{cleveref}
\usepackage[giveninits,date=year,sortcites]{biblatex}
\usepackage{tikz}
\usepackage{xspace}
\usepackage{listings}

\lstdefinestyle{cpp}{
    commentstyle=\color{black!50!white},
    basicstyle=\ttfamily\scriptsize,
    breakatwhitespace=true,
    breaklines=true,
    texcl=true,
}
\let\div\undefined
\DeclareMathOperator{\div}{div}
\DeclareMathOperator{\curl}{\mathbf{curl}}

\def\Hcurl{\bm{H}(\curl)}
\def\Hdiv{\bm{H}(\div)}

\def\Nedelec{N\'ed\'elec\xspace}

\def\T{\mathcal{T}}
\def\k{\kappa}
\def\khat{\widehat{\kappa}}
\def\Qp{\mathcal{Q}_p}

\def\Bd{B_{\rm 1D}}
\def\Dd{\partial_{\rm 1D}}

\def\Gin{\Gamma_{\text{in}}}
\def\Gout{\Gamma_{\text{out}}}
\def\Gbox{\Gamma_{\text{box}}}

\title[GPU-accelerated LOR preconditioning]{End-to-end GPU acceleration of low-order-refined preconditioning for high-order finite element discretizations}

\def\affila{1}
\def\affilb{2}

\author[Pazner]{Will Pazner$^{\affila,\affilb}$}
\address{$^\affila$Fariborz Maseeh Department of Mathematics and Statistics, Portland State University}
\author[Kolev]{Tzanio Kolev$^\affilb$}
\address{$^\affilb$Center for Applied Scientific Computing, Lawrence Livermore National Laboratory}
\author[Camier]{Jean--Sylvain Camier$^\affilb$}

\begin{document}

\begin{abstract}
   In this paper, we present algorithms and implementations for the end-to-end GPU acceleration of matrix-free low-order-refined preconditioning of high-order finite element problems.
   The methods described here allow for the construction of effective preconditioners for high-order problems with optimal memory usage and computational complexity.
   The preconditioners are based on the construction of a spectrally equivalent low-order discretization on a refined mesh, which is then amenable to, for example, algebraic multigrid preconditioning.
   The constants of equivalence are independent of mesh size and polynomial degree.
   For vector finite element problems in $\Hcurl$ and $\Hdiv$ (e.g.\ for electromagnetic or radiation diffusion problems) a specially constructed interpolation--histopolation basis is used to ensure fast convergence.
   Detailed performance studies are carried out to analyze the efficiency of the GPU algorithms.
   The kernel throughput of each of the main algorithmic components is measured, and the strong and weak parallel scalability of the methods is demonstrated.
   The different relative weighting and significance of the algorithmic components on GPUs and CPUs is discussed.
   Results on problems involving adaptively refined nonconforming meshes are shown, and the use of the preconditioners on a large-scale magnetic diffusion problem using all spaces of the finite element de Rham complex is illustrated.
\end{abstract}

\maketitle

\section{Introduction}

High-order finite element methods provide highly accurate solutions for problems with complex geometry using unstructured meshes, while simultaneously achieving efficiency and scalability on modern supercomputing platforms, in particular those with accelerator- and GPU-based architectures \cite{Kolev2021a,Brown2021,Hutchinson2016}.
While the efficient evaluation of high-order finite element operators using matrix-free methods and sum factorizations (cf.~\cite{Fischer2020,Kronbichler2019a,Orszag1980}) is well-studied, the solution of the resulting large linear systems remains challenging, in large part because the cost of assembling the system matrix of the high-order operator is prohibitive, both in terms of memory requirements and computation time.
In this work we describe the GPU acceleration and high-performance implementation of \textit{low-order-refined (LOR) preconditioning} for high-order finite element problems posed in $H^1$, $\Hcurl$, and $\Hdiv$ spaces.
These preconditioners, also known as ``FEM--SEM preconditioners'' (cf.~\cite{Orszag1980,Casarin1997,Canuto2010}), are based on the idea of constructing an auxiliary, spectrally equivalent low-order discretization, and then applying effective preconditioners (for instance, algebraic multigrid methods) constructed using the low-order matrix, directly to the high-order system.

There are a number of other approaches for the matrix-free preconditioning of high-order finite element problems.
One common approach is $p$-multigrid, which involves the construction of a hierarchy of $p$-coarsened spaces \cite{Helenbrook2006,Sundar2015}.
The coarsest space is typically a lowest-order space, which can be treated by means of a standard preconditioner for low-order finite element problems, for example algebraic multigrid methods.
On meshes that permit $h$-coarsening (for example, meshes that result from the successive refinement of an initial coarse mesh), geometric multigrid methods can also be used for matrix-free preconditioning.
These two approaches can be combined in a relatively flexible manner to obtain $hp$-coarsened hierarchies.
In all of these methods, a smoother is required at each level of the hierarchy; typically Jacobi or Chebyshev smoothing is performed (where the diagonal of the matrix is obtained without full matrix assembly using sum factorization, cf.\ \cite{Ronquist1987});
more sophisticated smoothers such as overlapping Schwarz may also be used \cite{Fischer1997}.
Combinations of these techniques with low-order-refined preconditioning is also possible, see \cite{Pazner2020a}.
Detailed performance comparisons of matrix-free methods for high-order discretizations are available in \cite{Kronbichler2018,Kronbichler2019,Phillips2022,Kolev2021}, among others.

In this work, we develop algorithms and software implementations to effectively leverage GPU-based computing architectures for the entire low-order-refined preconditioning algorithm.
This includes the parallel assembly of the auxiliary low-order system matrices, the construction and application of algebraic multigrid preconditioners, and the matrix-free application of the high-order operator in the context of a preconditioned conjugate gradient iteration.
For $\Hcurl$ and $\Hdiv$ problems, we use the AMS \cite{Kolev2009} and ADS \cite{Kolev2012} preconditioners, respectively, whose construction requires additional problem-specific inputs including mesh coordinates and discrete differential operators in matrix form; these inputs are also constructed on-device.
We provide comprehensive performance studies of these algorithms, measuring runtimes, kernel throughput, and strong and weak parallel scalability of all components of the solution algorithm.
Additionally, we compare the relative weightings of the algorithmic components on GPU and CPU, where our results show that the computational bottlenecks for GPU-accelerated solvers are significantly different than for CPU solvers.
In particular, solver setup is relatively more expensive (as a fraction of total time to solution) on GPU compared with CPU; however, the LOR matrix assembly algorithms proposed presently ensure that the assembly of the low-order-refined matrix is not a bottleneck.

The first work on LOR preconditioning dates to 1980, when Orszag proposed using second-order finite difference methods to precondition certain spectral collocation methods \cite{Orszag1980}.
Since then, there has been significant work extending this concept, making use of low-order finite element discretizations to precondition spectral and spectral element methods \cite{Canuto1994,Canuto2010,Fischer1997,Deville1985}.
The low-order discretization is often in turn preconditioned using multigrid, in particular algebraic or semi-structured multigrid methods \cite{Bello-Maldonado2019,Pazner2020a}.
Although these multigrid-based methods typically perform well, an attractive feature of LOR preconditioning is that \textit{any} effective preconditioner for the low-order system can be used to precondition the high-order system.
LOR preconditioning has been successfully applied to spectral element and high-order finite element discretizations of the incompressible Navier--Stokes equations \cite{Fischer2005,Franco2020}, and has been extended to other discretizations, including discontinuous Galerkin methods with $hp$-refinement \cite{Pazner2021b}.
Recently, spectrally equivalent LOR preconditioners for Maxwell and grad-div problems in $\Hcurl$ and $\Hdiv$ were developed \cite{Dohrmann2021a,Pazner2022}.

A novel aspect of this work is the use of a macro-element batching strategy for the assembly of the low-order-refined system matrices, and the associated algorithms and data structures.
There is significant existing work in the literature on GPU-accelerated algorithms for the matrix-free evaluation of high-order operators \cite{Ljungkvist2017,Abdelfattah2021,Franco2020};
these algorithms have been a primary focus of the CEED project targeting exascale architectures \cite{Kolev2021}.
Similarly, GPU-accelerated algebraic multigrid algorithms have been developed as part of the \textit{hypre} \cite{Falgout2002,Falgout2021} and AmgX \cite{Naumov2015} libraries, among others.
However, the efficient assembly of the low-order-refined matrices on GPUs is a topic that has not been addressed in these previous works.
The algorithms and GPU implementations described presently are shown to significantly outperform more generic matrix assembly approaches that do not take advantage of the macro-element-level regular structure and topology of the low-order-refined discretizations.

The structure of this paper is as follows.
In \Cref{sec:lor}, we give an overview of the LOR preconditioning approach for problems in $H^1$, $\Hcurl$, and $\Hdiv$, including on nonconforming meshes resulting from adaptive refinement.
In \Cref{sec:algorithms}, we describe the GPU algorithms for each of the algorithmic steps in the problem setup and the solution procedure.
In \Cref{sec:results}, we provide several detailed performance studies of the proposed methods, algorithms, and implementations.
The results include kernel-level throughput benchmarks, parallel scaling studies, and the illustration of these techniques on a more realistic and challenging large-scale electromagnetics problem.
Open-source code availability and a brief discussion of the software interface is provided in \Cref{sec:software}.
We end with conclusions in \Cref{sec:conclusions}.

\section{Low-order-refined preconditioning}
\label{sec:lor}

In this section, we give a brief overview of low-order-refined preconditioning for high-order finite element problems.
We begin by defining the low-order-refined mesh in \Cref{sec:lor-mesh}.
Low-order preconditioning for $H^1$-conforming discretizations of the Poisson problem is described in \Cref{sec:poisson}.
Preconditioning of problems in $\Hcurl$ and $\Hdiv$ are discussed in \Cref{sec:vector-fe}.

\subsection{The low-order-refined mesh} \label{sec:lor-mesh}
\begin{figure}
   \renewcommand{\tabcolsep}{10pt}
   \renewcommand{\arraystretch}{0}
   \begin{tabular}{ccc}
      \raisebox{-0.5\height}{\includegraphics[height=0.9in]{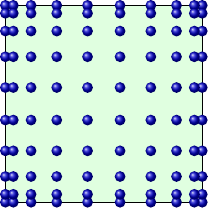}} &
      \raisebox{-0.5\height}{\includegraphics[height=1.25in]{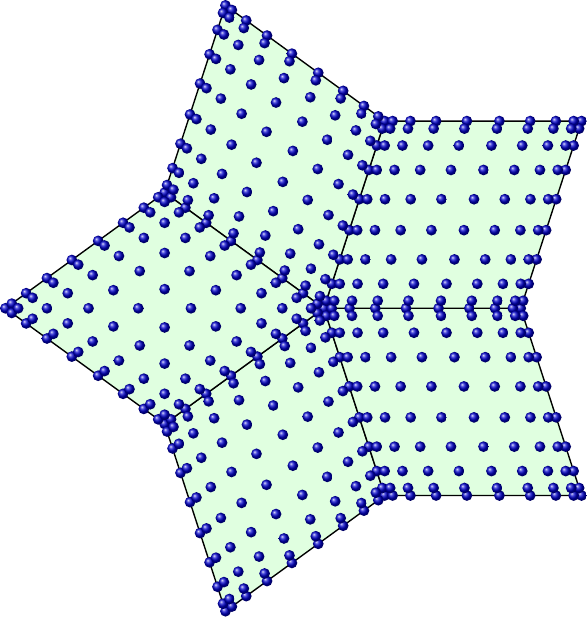}} &
      \raisebox{-0.5\height}{\includegraphics[height=1.25in]{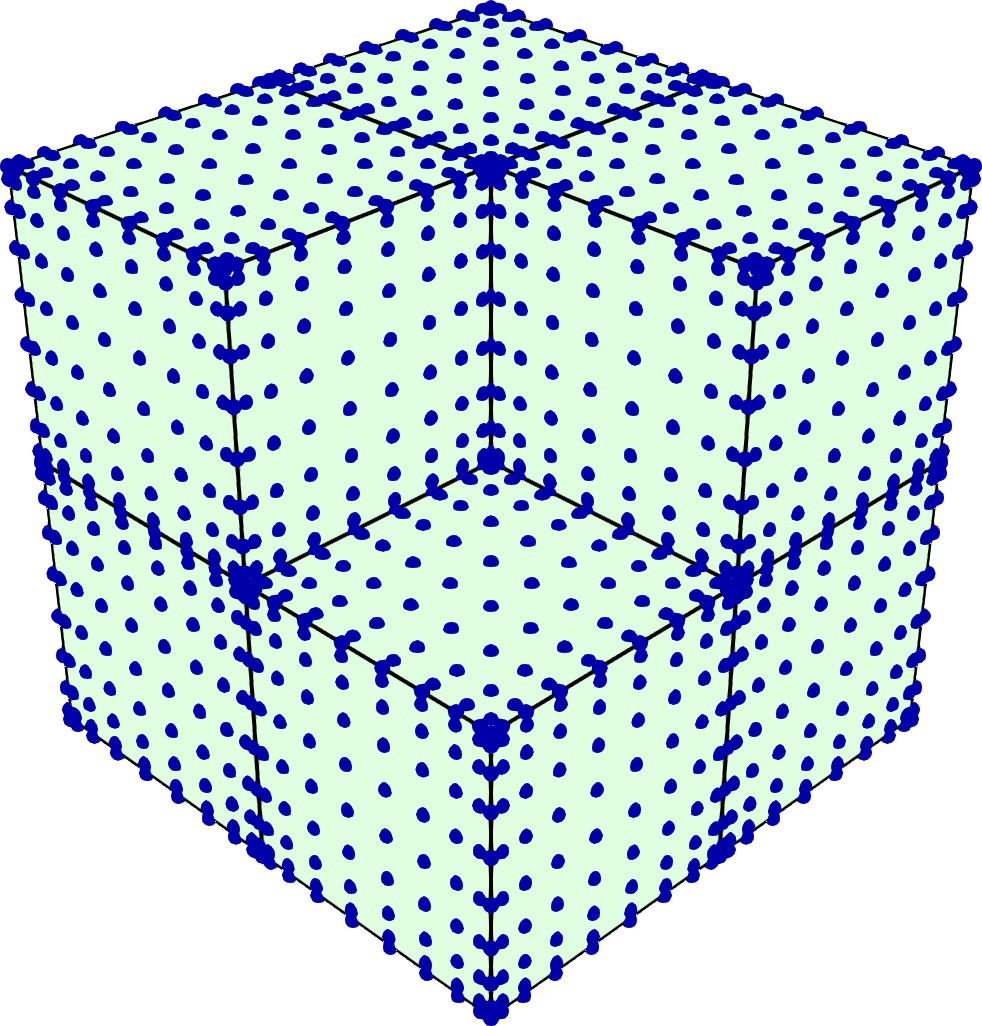}} \\[60pt]
      \raisebox{-0.5\height}{\includegraphics[height=0.9in]{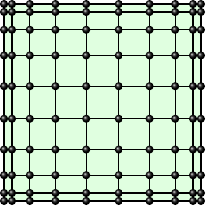}} &
      \raisebox{-0.5\height}{\includegraphics[height=1.25in]{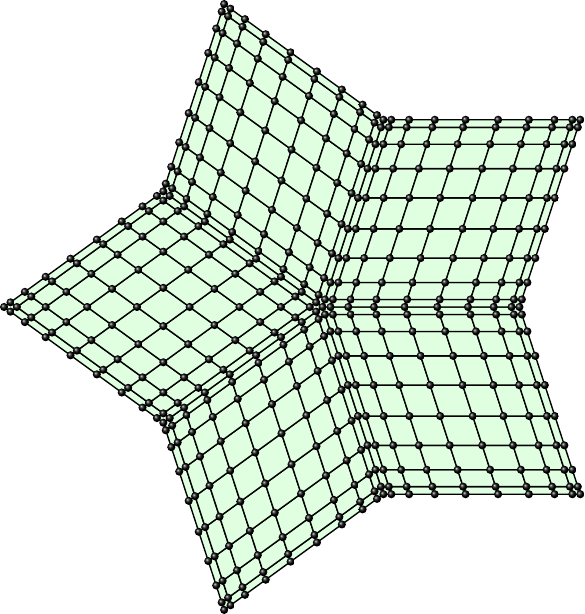}} &
      \raisebox{-0.5\height}{\includegraphics[height=1.25in]{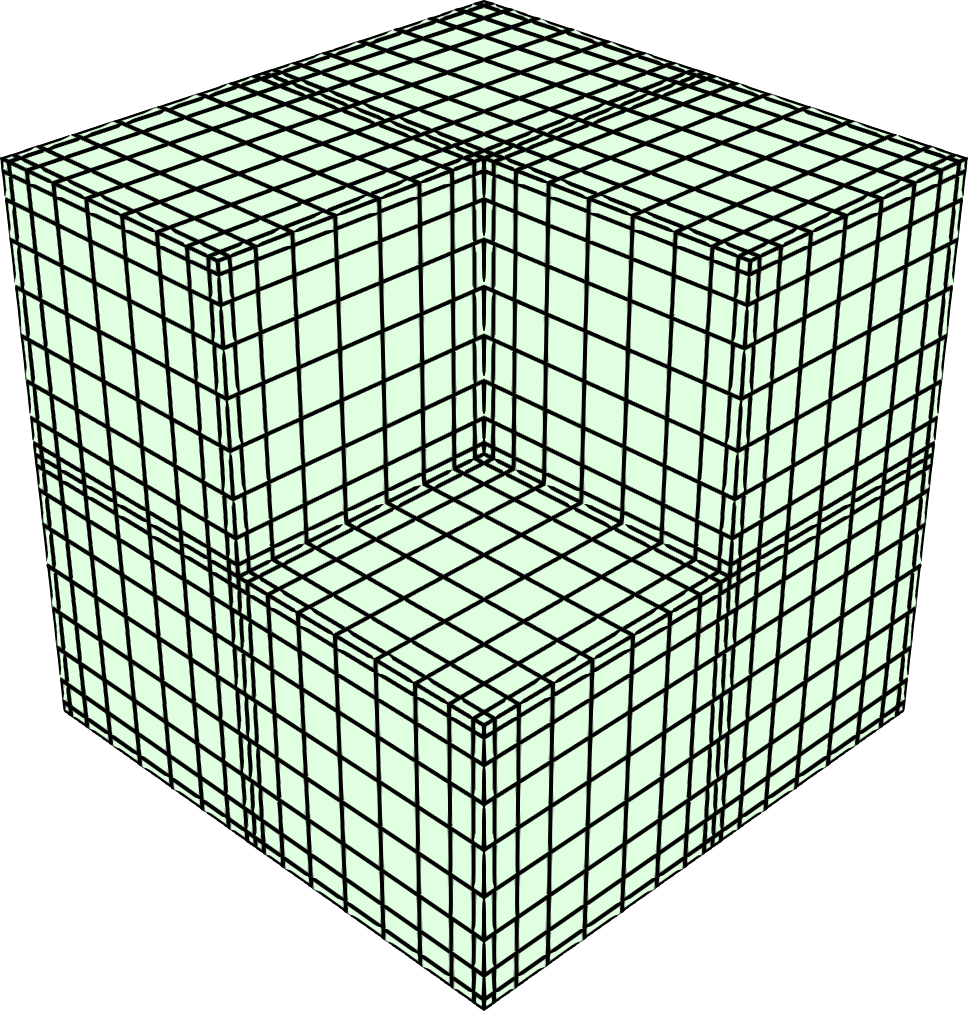}} \\
   \end{tabular}
   \caption{
      Top row: examples of high-order (coarse) meshes with $p=9$ Gauss--Lobatto nodes.
      Bottom row: corresponding low-order-refined meshes.
   }
   \label{fig:lor-meshes}
\end{figure}

Let $\Omega \subseteq \mathbb{R}^d$ denote the spatial domain, $d \in \{ 1, 2, 3 \}$.
The domain is discretized using a mesh $\T_p$ of tensor-product elements, such that each element $\k \in \T_p$ is given by the image of the reference element $\khat = [-1,1]^d$ under a (typically isoparametric) transformation, i.e.\ $\k = T_\k(\khat)$.
Let $\widehat{x}_i \in [-1,1]$ denote the $p+1$ Gauss--Lobatto points.
The $d$-fold Cartesian product of the points $\{ \widehat{x}_i \}$ is used as a set of Lagrange interpolation points in the reference element $\khat$.
These points are also used to define the \textit{low-order-refined mesh} $\T_h$.
Each element $\k \in \T_p$ is subdivided into $p^d$ topologically Cartesian subelements, whose vertices are given by the image of Gauss--Lobatto points under the element transformation $T_\k$.
Some examples of high-order (coarse) meshes $\T_p$ and the corresponding low-order-refined meshes $\T_h$ are given in \Cref{fig:lor-meshes}

\subsection{Poisson problem} \label{sec:poisson}

Consider the model Poisson problem
\begin{equation} \label{eq:poisson}
   \begin{aligned}
      -\Delta u &= f \qquad\text{in $\Omega$,}\\
      u &= g_D \qquad\text{on $\partial \Omega$.}
   \end{aligned}
\end{equation}
To discretize this problem, define the $H^1$-conforming degree-$p$ finite element space $V_p$,
\[
   V_p = \{ v \in H^1(\Omega) : v|_\k \circ T_\k \in \Qp(\khat) \},
\]
where $\Qp$ is the space of multivariate polynomials of at most degree $p$ in each variable.
A nodal basis for $\Qp$ is given by the Lagrange interpolating polynomials defined at the Cartesian product of the Gauss--Lobatto quadrature points, such that the degrees of freedom of the space $V_p$ are point values at the Gauss--Lobatto nodes.
The standard bilinear form $\mathcal{A}_{V_p} = (\nabla u, \nabla v)$ for $u, v \in V_p$ gives rise to the high-order stiffness matrix $A_{V_p}$.
In general, the total number of nonzeros in this matrix will scale like $\mathcal{O}(n_e p^{2d})$ where $n_e$ is the number of elements in the mesh, and the number of operations required to assemble this matrix ranges from $\mathcal{O}(n_e p^{2d + 1})$ to $\mathcal{O}(n_e p^{3d})$, depending on the algorithm used \cite{Melenk2001}.
This memory and computational costs are prohibitive for moderate to large values of $p$, and so instead a matrix-free approach is adopted, where the \textit{action} of $A_{V_p}$ is computed without constructing the matrix.
The action can be computed using sum factorization with optimal $\mathcal{O}(n_e p^d)$ memory usage and $\mathcal{O}(n_e p^{d+1})$ operations \cite{Orszag1980};
this matrix-free operator evaluation approach is the key component of the high-performance finite element libraries software library libCEED \cite{Brown2021} and also provides the foundation for MFEM's \textit{partial assembly} algorithm \cite{Anderson2020}.

Because of the prohibitive assembly costs, the matrix $A_{V_p}$ is unavailable for preconditioner construction.
Instead, we construct an auxiliary low-order discretization matrix that is spectrally equivalent to the high-order matrix.
The low-order-refined space $V_h$ is defined to be the $p=1$ $H^1$-conforming space on the low-order-refined mesh $\T_h$, whose construction is described in \Cref{sec:lor-mesh}.
The degrees of freedom of $V_h$ coincide with the degrees of freedom of the high-order space $V_p$.
In both cases, the degrees of freedom represent point values at the Gauss--Lobatto points;
in the high-order case, the associated basis functions are degree-$p$ polynomials defined on the elements of the coarse mesh $\T_p$, and in the low-order case, the associated basis functions are the standard ``hat functions'' defined on the refined mesh $\T_h$.
Given the low-order space $V_h$, it is possible to assemble the low-order-refined stiffness matrix $A_{V_h}$.
We briefly summarize some important properties of the matrices $A_{V_h}$ and $A_{V_p}$.

\begin{itemize}
   \item Since the high-order and low-order-refined degrees of freedom coincide, $A_{V_h}$ and $A_{V_p}$ are matrices of the same size.
   \item The low-order matrix $A_{V_h}$ has $\mathcal{O}(1)$ nonzeros per row. On the other hand, the high-order matrix $A_{V_p}$ has $\mathcal{O}(p^{2d})$ nonzeros per row.
   \item $A_{V_h}$ and $A_{V_p}$ are spectrally equivalent, independent of $p$. This spectral equivalence is often known as \textit{FEM--SEM equivalence}, and was proven in \cite{Canuto1994} and \cite{Parter1995}.
\end{itemize}

Since $A_{V_h}$ and $A_{V_p}$ are spectrally equivalent, any uniform preconditioner $B_h$ of $A_{V_h}$ (i.e.\ such that the condition number of $B_h A_{V_h}$ is bounded independently of problem size and other discretization parameters) will also be an effective preconditioner for $A_{V_p}$.
In this work, we take $B_h$ to be one V-cycle of algebraic multigrid (AMG) constructed using the assembled matrix $A_{V_h}$.
Other choices for $B_h$ are possible, including geometric or semi-structured multigrid, domain decomposition, or even sparse direct solvers, but in this work we restrict ourselves to AMG methods.
Note that the same considerations regarding sparsity also apply to the high-order and low-order-refined mass matrices (denoted $M_{V_p}$ and $M_{V_h}$, respectively).
These mass matrices are spectrally equivalent (with constants of equivalence independent of polynomial degree $p$), and so low-order-refined preconditioning can be also applied to problems involving non-negative linear combinations of the mass and stiffness matrices.

\subsection{Vector finite elements} \label{sec:vector-fe}

While the approach described above can be used to obtain spectrally equivalent low-order-refined preconditioners using nodal $H^1$ discretizations for Poisson problems, the situation is more delicate for Maxwell and grad-div problems in $\Hcurl$ and $\Hdiv$.
In these spaces, nodal bases using Gauss--Lobatto or Gauss--Legendre points do not give rise to spectrally equivalent low-order discretizations, and as a result, the condition number of the preconditioned system grows rapidly, leading to large iteration counts.

Instead, an \textit{interpolation--histopolation} basis for the high-order $\Hcurl$ and $\Hdiv$ spaces must be used \cite{Dohrmann2021a,Pazner2022}.
The resulting basis functions are piecewise polynomials that take on prescribed mean values over subcell edges (in the case of $\Hcurl)$ or subcell faces (in the case of $\Hdiv$).
These basis functions were introduced in \cite{Kreeft2011} in the context of mimetic methods.
When this basis is used, spectral equivalence of the high-order and low-order-refined stiffness matrices (and mass matrices) are recovered for vector finite element spaces.
The construction of low-order-refined preconditioners for this case is discussed at length in \cite{Pazner2022}.
We note that the lowest-order case of the interpolation--histopolation bases reduces exactly to the standard lowest-order \Nedelec and Raviart--Thomas elements.
As in the case of the Poisson problem, any effective preconditioner constructed using the low-order-refined system matrix can be used to effectively precondition the high-order system.
However, even in the low-order case, the construction of preconditioners for these problems is more challenging.
Typically, multigrid methods with specialized smoothers \cite{Arnold2000} or auxiliary space methods \cite{Hiptmair2007} are required to achieve good convergence.
Analogous to the use of algebraic multigrid for the Poisson problem, in this work we focus on the use of auxiliary space AMG methods for these problems;
in particular, we use the AMG-based algebraic Maxwell solver (AMS) for $\Hcurl$ problems \cite{Kolev2009}, and the auxiliary divergence solver (ADS) for $\Hdiv$ problems \cite{Kolev2012}.
These solvers are based on combining auxiliary space methods with algebraic multigrid, and are designed to work as black-box methods, requiring relatively little discretization information beyond the assembled system matrix.
The AMS $\Hcurl$ solver requires two additional inputs: a vector of mesh vertex coordinates (used to construct the interpolation operator mapping the vector $H^1$ to the \Nedelec space; this interpolation matrix can also be provided directly), and a discrete gradient matrix.
In addition to the discrete gradient and interpolation, the ADS $\Hdiv$ solver additionally requires a discrete curl matrix, mapping from $\Hcurl$ into $\Hdiv$.
The construction of these discrete matrices is also discussed in the following sections.

\subsection{Adaptive mesh refinement and hanging node constraints} \label{sec:amr}

\begin{figure}
   \sf \footnotesize
   \begin{tabular}{cc@{\qquad}cc}
      \includegraphics[width=0.2\linewidth]{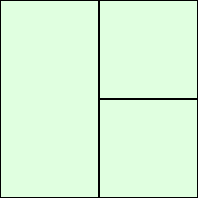} &
      \includegraphics[width=0.2\linewidth]{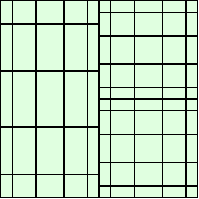} &
      \includegraphics[width=0.2\linewidth]{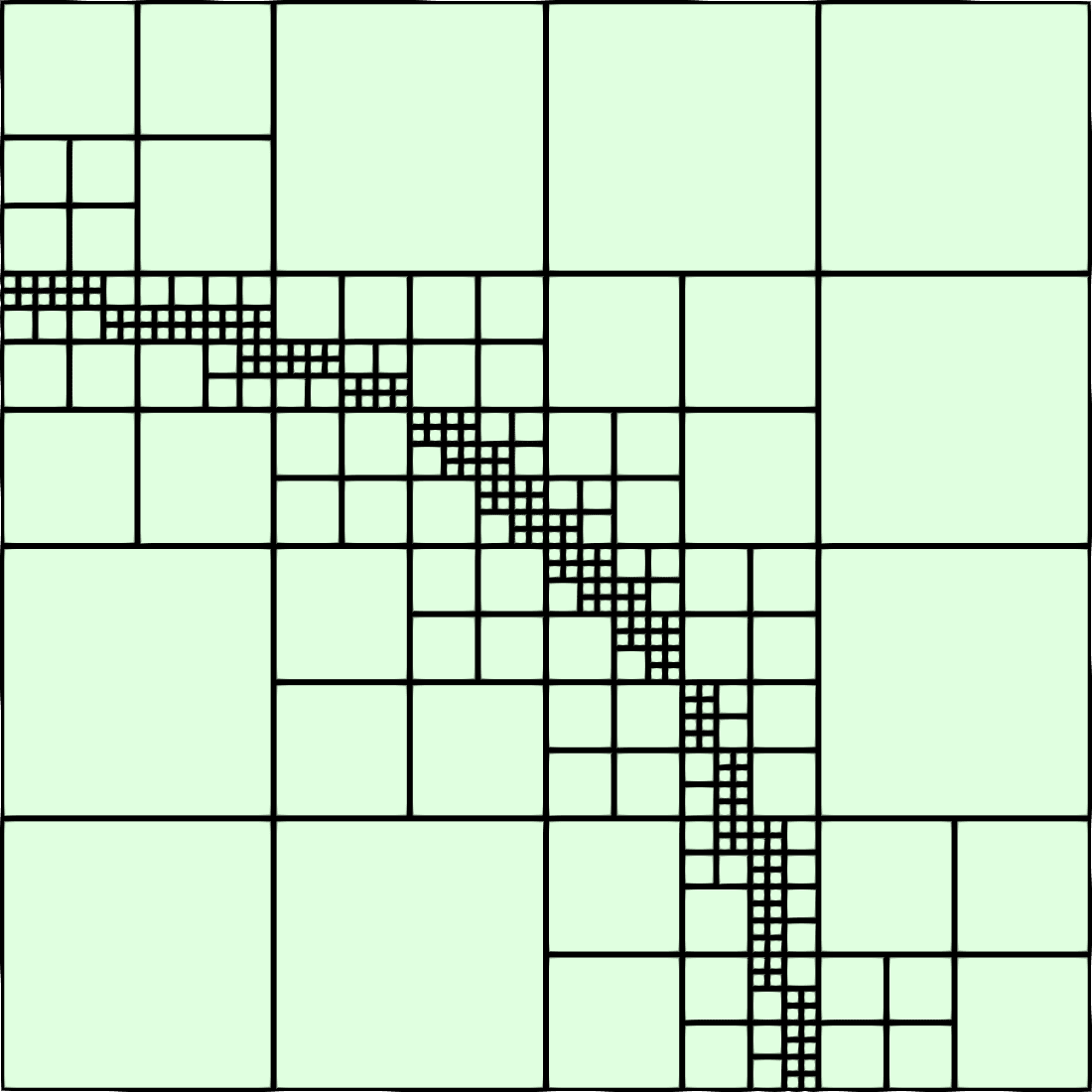} &
      \includegraphics[width=0.2\linewidth]{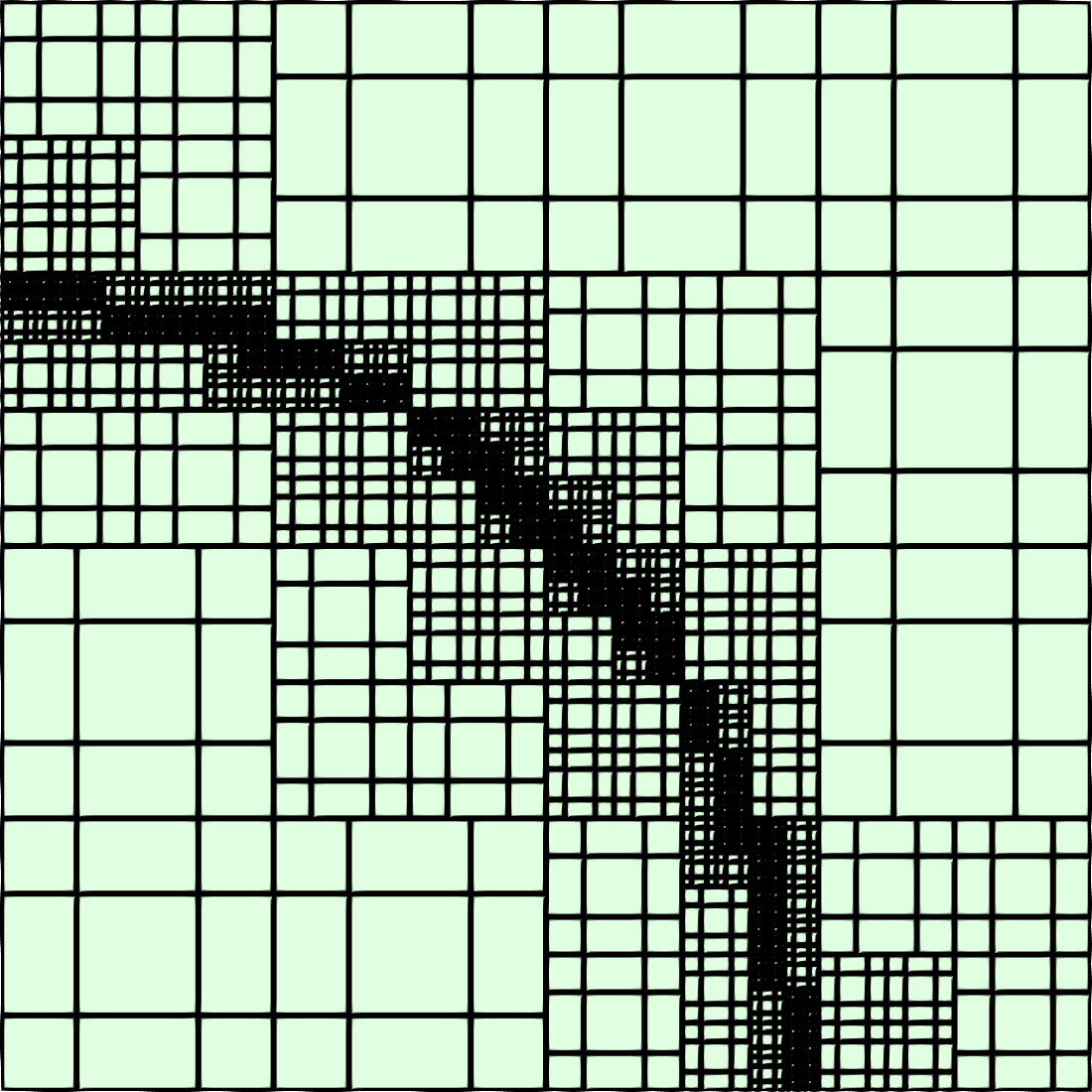} \\
      (a) & (b) & (c) & (d) \\[10pt]
      \multicolumn{2}{p{0.4\linewidth}}{(a) Mesh with three elements and one hanging node. \newline (b) Low-order-refined mesh corresponding to $p=5$.} &
      \multicolumn{2}{@{}p{0.4\linewidth}}{(c) Adaptively refined mesh. \newline (d) Low-order-refined mesh corresponding to $p=3$.}
   \end{tabular}
   \caption{
      Illustration of nonconforming meshes with hanging nodes and their low-order-refined counterparts.
   }
   \label{fig:amr-meshes}
\end{figure}

In the case of meshes with hanging nodes (e.g.\ those resulting from nonconforming adaptive mesh refinement), the construction of the appropriate low-order-refined discretization is less obvious.
In this case, the refinement procedure described in \Cref{sec:lor-mesh} to obtain the low-order-refined mesh results in non-matching interfaces that do not directly lend themselves to the construction of spectrally equivalent low-order discretizations, see \Cref{fig:amr-meshes}.

In \cite{Pazner2021b}, an approach for constructing low-order-refined preconditioners for discretizations posed on meshes with hanging nodes based on variational restriction was proposed.
In this approach, the high-order system is written as
\begin{equation} \label{eq:Ap-amr}
   A_p = \Lambda^T \widehat{A}_p \Lambda,
\end{equation}
where $\Lambda$ is the matrix that enforces the constraints at element interfaces, and $\widehat{A}_p$ is a block-diagonal matrix acting on vectors of unconstrained (duplicated) degrees of freedom with element matrices $\widehat{A}_{p,i}$ on the diagonal.
The matrix $\Lambda$ is often known as an \textit{assembly matrix};
in the case where the mesh contains no hanging nodes (and without $p$-refinement), the matrix $\Lambda$ is a boolean matrix encoding the local (elementwise) to global DOF mapping.
When the mesh contains hanging nodes, constraints must be enforced at the nonconforming interfaces, resulting in a more complicated $\Lambda$ matrix;
see also \cite{Cerveny2019} for more details.

The low-order-refined counterpart $\widehat{A}_{h,i}$ of each element matrix $\widehat{A}_{p,i}$ can be assembled to obtain a block-diagonal matrix $\widehat{A}_h$.
The diagonal blocks of $\widehat{A}_h$ are sparse, whereas the blocks of $\widehat{A}_p$ are in general dense.
Then, a low-order-refined preconditioner can be formed by
\begin{equation} \label{eq:Ah-amr}
   A_h = \Lambda^T \widehat{A}_h \Lambda,
\end{equation}
where the same matrix $\Lambda$ is used in \eqref{eq:Ap-amr} and \eqref{eq:Ah-amr}.
In the case of conforming meshes, the matrix $A_h$ defined by \eqref{eq:Ah-amr} is identical to that obtained through the process described in \Cref{sec:poisson}.
When the mesh is nonconforming, a simple variational argument shows that $A_h$ is spectrally equivalent to $A_p$, and therefore can be used as a preconditioner.
The implementation of such preconditioners requires only the construction of the constraint matrix $\Lambda$, and the computation of the triple product in \eqref{eq:Ah-amr}.
Then, AMG or other algebraic preconditioners can be constructed using the assembled matrix $A_h$, and used to precondition the high-order system defined by $A_p$.

\section{GPU algorithms} \label{sec:algorithms}

The LOR-based solution procedure can broadly be divided into two phases.
The \textit{setup phase}, performed before the beginning of the conjugate gradient iteration, consists of problem setup, matrix assembly, and construction of the algebraic multigrid preconditioner.
The \textit{solve phase} consists of the preconditioned conjugate gradient iteration, which at every step requires the matrix-free application of the high-order operator and the application of the algebraic multigrid V-cycle.
In this section we describe the approach for GPU-accelerated algorithms for each of these components.
We summarize the algorithmic steps that make up the solution algorithm:
\begin{enumerate}[label*=S\arabic*.,ref=S\arabic*]
   \item Problem setup:
   \begin{enumerate}[label*=\arabic*.,ref=S\arabic{enumi}.\arabic*]
      \item \label{item:pa-setup} High-order operator setup;
      \item \label{item:lor-assembly} Low-order-refined matrix assembly;
      \item \label{item:amg-setup} Algebraic multigrid setup.
   \end{enumerate}
   \item Preconditioned conjugate gradient, including at every iteration:
   \begin{enumerate}[label*=\arabic*.,ref=S\arabic{enumi}.\arabic*]
      \item \label{item:pa-apply} High-order operator evaluation;
      \item \label{item:amg-solve} Algebraic multigrid V-cycle.
   \end{enumerate}
\end{enumerate}

Steps \ref{item:pa-setup} and \ref{item:pa-apply} are the main algorithmic components for the fast evaluation of matrix-free high-order operators, as described in detail in \cite{Anderson2020,Kolev2021,Abdelfattah2021,Brown2021,Kronbichler2019a}, among others;
efficient GPU-accelerated algorithms and implementations for these operations have been extensively discussed in the literature.
Likewise, steps \ref{item:amg-setup} and \ref{item:amg-solve} make up the two phases of algebraic multigrid methods, and efficient GPU strategies have been discussed in \cite{Haase2010,Falgout2021,Bell2012,Naumov2015}.
Although we will briefly summarize the important aspects of these algorithms in the following sections, the primary contribution of this work is the development of efficient, GPU-suitable algorithms Step \ref{item:lor-assembly}.

% The following subsections aren't automatically numbered. Instead, they are given section titles of
% the form "Step SX.Y.", where "SX.Y" is the step number from the above list. When we
% cross-reference such a section, we want the reference to appear as "step X.Y" or "Step X.Y"
% depending on capitalization, so we use \steplabel defined below instead of label.
%
% Argument 1 - the name of the step item and of the section, e.g. for "item:pa-setup", this should
%              "pa-setup", and the resulting section will have label "sec:pa-setup".
\newcommand{\steplabel}[1]{
   \crefformat{#1-step}{step ##2\ref*{item:#1}##3}
   \Crefformat{#1-step}{Step ##2\ref*{item:#1}##3}
   \label[#1-step]{sec:#1}
}

\subsection*{Step \ref{item:pa-setup}. High-order operator setup}
\steplabel{pa-setup}

\begin{figure}
   \includegraphics{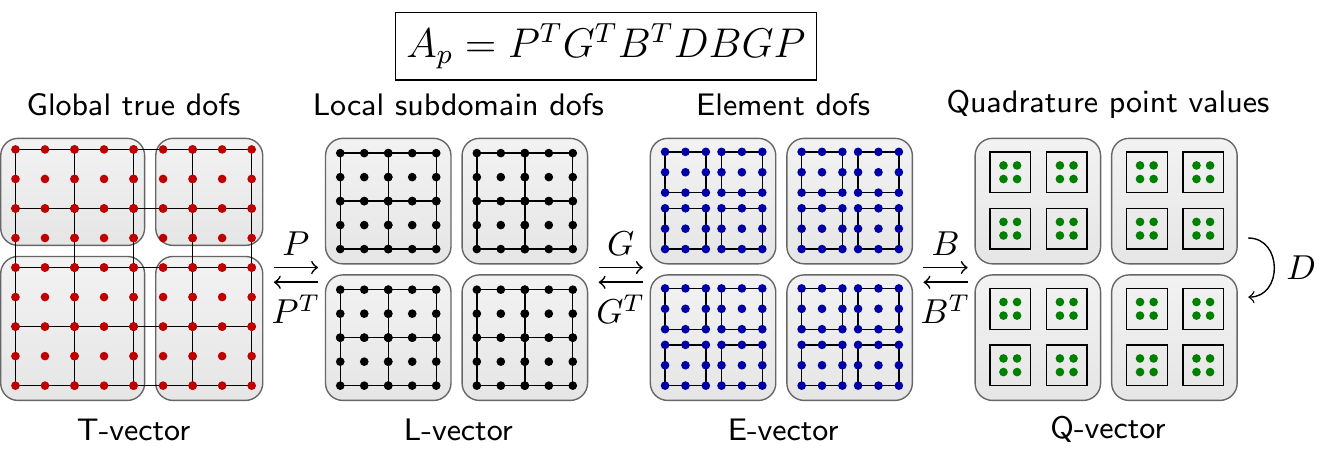}
   \caption{Schematic of the high-order operator decomposition used in the partial assembly approach in the MFEM and libCEED software libraries.}
   \label{fig:libceed}
\end{figure}

To achieve good performance with high-order discretizations, especially on GPUs, the high-order operator is evaluated matrix-free; the discretization matrix associated with the operator is never assembled.
Instead, we make use of a matrix-free approach termed \textit{partial assembly}.
The high-order operator, denoted $A_p$, is decomposed as a nested product of matrices,
\begin{equation} \label{eq:op-decomp}
   A_p = P^T G^T B^T D B G P,
\end{equation}
see also \Cref{fig:libceed}.
This decomposition is used heavily in libCEED \cite{Brown2021}, MFEM \cite{Anderson2020}, and associated software.
In the above, $P$ denotes the parallel prolongation operator (resulting from the parallel decomposition of the spatial domain) mapping global parallel vectors (referred to as T-vectors) to processor-local vectors (L-vectors), $G$ denotes the restriction from an L-vector of conforming degrees of freedom, to a (``broken'') vector of element-wise degrees of freedom (referred to as an E-vector), $B$ is the basis operator, that evaluates the shape functions (and their derivatives) at quadrature points, and $D$ is a pointwise (diagonal or block-diagonal) matrix that contains the geometric factors of the mesh and point values of problem-specific variable coefficients.

In the partial assembly framework, the product in \eqref{eq:op-decomp} is not formed, and instead the individual factors are applied sequentially.
The factors are themselves typically not stored as sparse matrices.
Instead, the matrices $P$ and $G$ make use of degree of freedom index mappings that are precomputed and stored; these mappings depend only on the mesh and finite element space, and not on the specific operator or physics of the problem.
In the case of nonconforming adaptive mesh refinement, the $P$ matrix also incorporates the AMR constraints, and is represented as a parallel CSR matrix; when the mesh is conforming, $P$ is represented matrix-free.
The basis operator $B$ is identical between all elements of the same type (geometry and polynomial degree).
In this work we consider meshes consisting of tensor-product elements with constant polynomial degree, in which case one element-local $B$ matrix can be stored and used for all elements in the mesh.
Additionally, the tensor-product structure of quadrilateral and hexahedral elements allows for $B$ to be expressed as a Kronecker product, requiring only the storage of small one-dimensional factors.
The approach is closely related to sum factorization, which has been widely adopted in the areas of spectral and spectral element methods \cite{Orszag1980}.
Finally, the $D$ matrix consists of either scalar values or small matrices ($d \times d$, where $d$ is the spatial dimension) at each quadrature point.
These values are precomputed and stored as part of the operator setup.

Partial assembly refers to the precomputation of the degree of freedom index mappings required for $P$ and $G$, the one-dimensional interpolation and derivative matrices required for $B$, and the construction of the $D$ matrix.
Given the geometric factors at quadrature points, all of these precomputations can be performed in constant time per degree of freedom (that is, linear time in the problem size, regardless of the polynomial degree of the finite element space).
The evaluation of the geometric factors itself can be performed using sum factorization, resulting in a scaling of $\mathcal{O}(p_g^{d+1})$, where $p_g$ is the polynomial degree associated with the element transformation mapping ($p_g = 1$ for straight-sided meshes, $p_g = p$ for isoparametric curved meshes).

\subsection*{Step \ref{item:lor-assembly}. Low-order-refined matrix assembly}
\steplabel{lor-assembly}

Although the action of the high-order operator is computed matrix-free, in order to construct the low-order-refined algebraic multigrid preconditioners, it is required to assemble the system matrix associated with the low-order-refined discretization.
This matrix has the same dimensions as that of the high-order operator, but is significantly sparser:
while number of nonzeros per row of the high-order matrix scales like $\mathcal{O}(p^d)$, the number of nonzeros per row in the low-order-refined matrix remains bounded, independent of $p$.
As a result, the assembly cost and memory usage of the low-order-refined matrix is linear in the problem size, which is optimal.

While the asymptotic complexity of the matrix assembly is known to be optimal, the achieved throughput of these operations on GPU-based architectures depends greatly on the structure of the assembly algorithms, the choice of data structures, and the threading strategies.
In particular, it has been widely shown that high-order finite element operations often achieve greater throughput for similarly sized problems when compared with their low-order counterparts, despite requiring more arithmetic operations \cite{Kolev2021,Franco2020}.
In large part, this is because typical finite element operations are memory-bound, and the cost of performing arithmetic operations is negligible compared with the cost of memory transfer and access.
High-order operators lend themselves to more structured and efficient memory access patterns than low-order operators, leading to greater performance.
This presents a challenge for the assembly of low-order-refined matrices, which inherently use lowest-order finite elements.

\begin{figure}
   \begin{tikzpicture}
      \node at (0,0) {
         \includegraphics[width=2in]{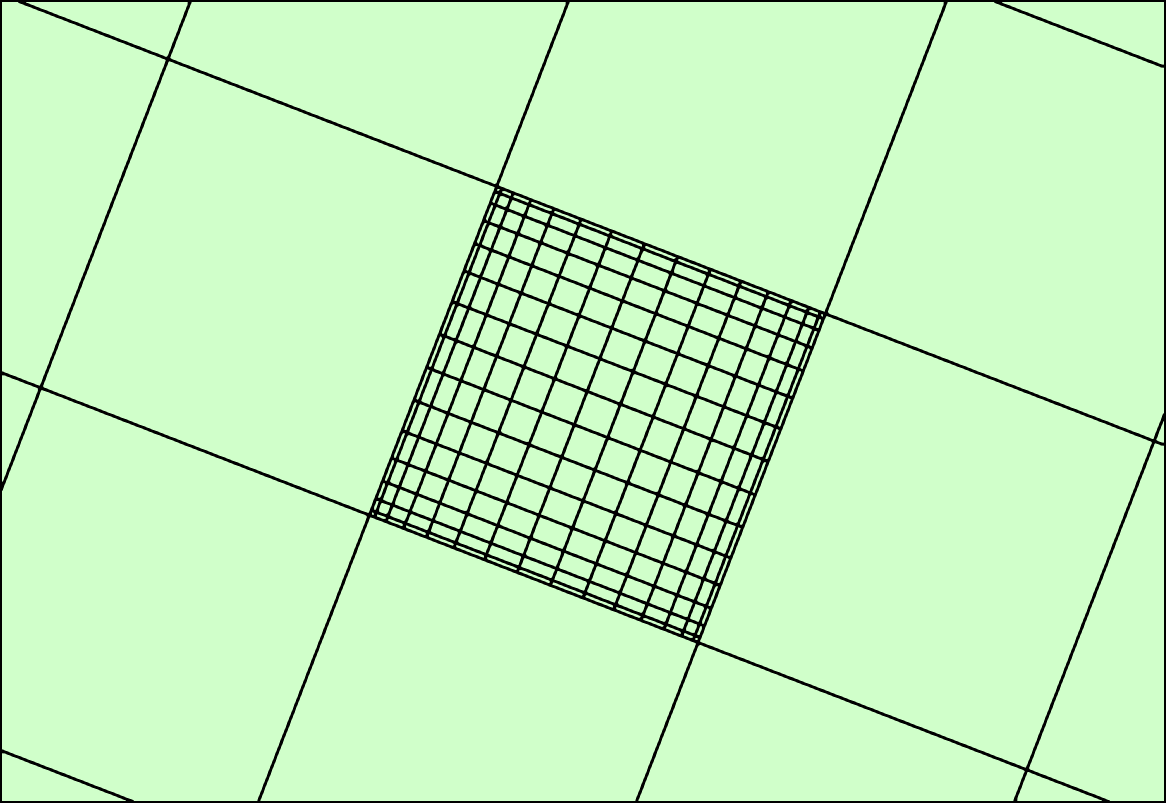}
      };
      \node at (-1.4,0.5) {\large $\T_p$};
      \draw[-stealth] (3.5,0.5) node[anchor=west] {Macro element in $\T_h$} to[in=0,out=180] (0.8,-0.5);
   \end{tikzpicture}
   \caption{An example of a high-order mesh $\T_p$ with a macro element of the LOR mesh $\T_h$ overlaid.}
   \label{fig:macro-element}
\end{figure}

The approach taken in this work is to structure the assembly of the low-order-refined matrices around \textit{macro elements}.
In the context of low-order-refined methods, each element of the coarse, high-order mesh $\T_p$ is refined into $p^d$ subelements to obtain the refined mesh $\T_h$.
The collection of $p^d$ refined elements in $\T_h$ corresponding to a single high-order element is referred to as a macro element.
An illustration of a macro element is shown in \Cref{fig:macro-element}.
The advantage of working at the level of macro elements is that within each macro element, the local mesh topology is that of a structured Cartesian mesh.
Furthermore, this local mesh topology is identical between all macro elements, repeated across the mesh.
This structure allows for application of algorithmic components similar to those of high-order finite elements to the case of low-order-refined assembly.

The overall matrix assembly algorithm proceeds according to the following steps, which are described in detail in the following sections.
\begin{enumerate}[label*=A\arabic*.,ref=A\arabic*]
   \item \label{item:asm-macro} Assembly of sparse matrices for each macro element.
   \item \label{item:asm-local} Assembly of processor-local sparse matrix.
   \item \label{item:asm-global} Assembly of global (parallel) sparse matrix.
   \item \label{item:asm-bcs} Elimination of boundary conditions.
\end{enumerate}

\subsubsection*{Step \ref{item:asm-macro}. Macro-element sparse matrix assembly}

Traditional algorithms for finite element assembly proceed by constructing small, independent element-local matrices (in ``unassembled form''), and then placing them in the system matrix.
This can be expressed algebraically as
\[
   A = \Lambda^T \widehat{A} \Lambda,
\]
where $\widehat{A}$ denotes a block-diagonal matrix, where the $i$th diagonal block $\widehat{A}_i$ consists of the local matrix of the $i$th element, and $\Lambda$ is the assembly matrix introduced in \Cref{sec:amr}.
In the case of low-order-refined methods, it is inefficient to work at the level of individual elements, because this would result in fully unstructured memory access, and additional duplication of matrix entries corresponding to shared degrees of freedom.
Instead, we construct blocks $\widehat{A}_i$ that are local to each \textit{macro element}.
To take advantage of the sparsity of the LOR discretization, the macro-element matrices are stored in sparse rather than dense format.
Because of the uniform Cartesian structure within each macro element, the sparsity pattern for each block $\widehat{A}_i$ is identical, and so the sparse matrix graph need only be stored once for the entire problem.
Each block $\widehat{A}_i$ is stored in a modified CSR format.
In this format, the row array (denoted $\widehat{\texttt{I}}$) is implicit, since as a result of the Cartesian structure, most rows have the same number of nonzeros (for example, the local matrices of the low-order-refined diffusion operator have 9 nonzeros per row in 2D and 27 nonzeros per row in 3D, with the exception of rows corresponding to vertices that lie on the boundary of the macro element, which have fewer nonzeros).
We will use \texttt{nnz\_per\_row} to denote this bound on the number of nonzeros per row.
The column pointer array (denoted $\widehat{\texttt{J}}$) is computed once and shared between all macro elements.
It has shape $\texttt{nnz\_per\_row} \times \texttt{ndof\_per\_el}$, where $\texttt{ndof\_per\_el}$ is the number of degrees of freedom per macro element.
If a row has fewer than \texttt{nnz\_per\_row} nonzeros (i.e.\ the row corresponds to a degree of freedom lying on the macro element boundary), then $\widehat{\texttt{J}}$ is padded with -1, which represents an invalid index.
The array of CSR nonzeros (denoted $\widehat{\texttt{A}}$) is stored as a tensor of shape $\texttt{nnz\_per\_row} \times \texttt{ndof\_per\_el} \times \texttt{nel}$, where $\texttt{nel}$ is the number of macro elements in the mesh $\T_h$ (that is, the number of coarse elements in the mesh $\T_p$).

The main work of Step \ref{item:asm-macro} is to fill in the entries of the $\widehat{\texttt{A}}$ array, since $\widehat{\texttt{I}}$ is omitted, and $\widehat{\texttt{J}}$ is independent of the number of elements in the mesh.
This procedure is carried out on the GPU using one block of threads per macro element.
In 2D, the block of threads is of size $p \times p$, corresponding to the $p^2$ subelements in each macro element.
In 3D, the block of threads is of size $p \times p \times \widetilde{p}$, where the last dimension $\widetilde{p}$ may be lowered to avoid register spilling.
The matrix associated with each subelement is assembled by one thread, and then placed in the $\widehat{\texttt{A}}$ array, which is in global memory.
Since we are presently concerned with symmetric problems, only one of the upper or lower triangular parts of the local matrix need be assembled.
The placement into the $\widehat{\texttt{A}}$ array requires atomic operations, because subelements sharing a common degree of freedom may be assembled by different threads.

In order to assemble the subelement matrices, geometric factors are required at the subelement vertices.
In the completely unstructured approach, this would first require the computation of the mesh coordinates at each low-order-refined vertex in ``broken'' element-wise format (i.e.\ as an LOR E-vector), resulting in unnecessary duplication and inefficient access patterns.
Instead, in the macro element approach, the mesh coordinates are represented as high-order E-vectors, where all the coordinates corresponding to a macro element are stored contiguously, with no duplication at the subelement level.
This is more efficient both in terms of memory usage and access patterns, and also allows the reuse of the high-order $G$ operator, as described in Step \ref{item:pa-setup}.

\subsubsection*{Step \ref{item:asm-local}. Processor-local assembly}
\steplabel{asm-local}

Once the macro element sparse matrices $\widehat{A}_i$ have been constructed, the next step is to assemble a processor local matrix in CSR format.
To determine the number of nonzeros for each degree of freedom, the nonzeros indices in each macro element sparse block are added, taking care not to double-count nonzeros indices corresponding to degrees of freedom that lie on macro element interfaces.
This deduplication is achieved by adding only those nonzeros indices that correspond to the minimal macro element index among all macro elements containing the specified degree of freedom.
This is performed using $\texttt{ndof\_per\_el} \times \texttt{nel}$ threads, and as before, atomic operations must be used to avoid conflicts.
A scan operation is then performed to construct the processor-local \texttt{I} array.

Once \texttt{I} has been constructed, \texttt{J} and \texttt{A} are computed using the sparse blocks $\widehat{A}_i$.
Again, $\texttt{ndof\_per\_el} \times \texttt{nel}$ threads are used, and each thread loops over the nonzeros of the corresponding row of $\widehat{A}_i$.
The responsibility to add nonzero values to the CSR arrays is left to the thread belonging to the macro element with the minimal index among all macro elements with the given nonzero index.
The column pointer array $\widehat{\texttt{J}}$ is used to identify the column index for each nonzero.
For vector finite elements in $\Hcurl$ and $\Hdiv$, the orientation of the vector-valued basis functions is encoded in the sign of the degrees of freedom.
At the macro element level, because of the Cartesian structure, lexicographic ordering and orientation is used.
At the processor-local level, the orientation is inherited from that of the mesh.
This sparse matrix constructor procedure is also responsible for ensuring consistent orientations during assembly.

\subsubsection*{Step \ref{item:asm-global}. Global matrix assembly}

When running on a single MPI rank (i.e.\ on a single GPU), this step can be skipped entirely.
When running in parallel, the construction of the global (parallel) CSR matrix is required.
In this format, the degrees of freedom are partitioned by MPI rank, and each rank owns the matrix rows associated with its degrees of freedom.
Each rank has a diagonal block, which contains the nonzeros for which both the row and column indices are owned by itself, and an off-diagonal block, which contains the nonzeros for which the column indices are owned by another rank.
The construction of this parallel CSR matrix is performed using a triple product $P^T A P$.
Because this operation (or $RAP$ more generally) is important for the construction of coarse operators in algebraic multigrid, the \textit{hypre} library provides optimized GPU kernels for computing triple products.
This operation can further be optimized using the knowledge that in the case of conforming meshes, the $P$ matrix is boolean.
For details on the algorithms used for this operation, see \cite{Falgout2021}.

\subsubsection*{Step \ref{item:asm-bcs}. Elimination of boundary conditions}

Once the global matrix has been formed, the matrix must be modified to take into account essential boundary conditions.
The rows and columns corresponding to each essential degree of freedom must be eliminated, and the corresponding diagonal entry is typically replaced by the value 1.
The elimination of columns in parallel requires communication, because a column corresponding to a degree of freedom owned by one rank may have nonzeros in rows owned by other ranks.
The first step of the elimination procedure is to communicate to each rank which columns it must eliminate.
This communication is performed using non-blocking, device-aware MPI, whenever available.
The column indices to eliminate are obtained in the form of a marker array, i.e.\ a boolean array with 1 in the indices that must be eliminated, and 0 elsewhere.

Once the communication has begun, the rows and columns in the diagonal block, and the rows in the off-diagonal block can be eliminated, thus allowing for the overlap of communication and computation.
The elimination is embarrassingly parallel, and is threaded over the number of essential degrees of freedom.
After this procedure is complete, we wait for the communication to finish, and finally eliminate the columns in the off-diagonal block.
Since this data is available in the format of a marker array, each nonzero in the off-diagonal block is simply scaled by one or zero depending on the marker value, allowing parallelization over the number of rows in the off-diagonal.

\subsubsection*{Auxiliary computations}

The above sections describe the main algorithmic components of the low-order-refined matrix assembly.
In the case of vector finite element problems in $\Hcurl$ and $\Hdiv$, certain additional discretization information is required by the AMS and ADS algebraic solvers.
In particular, both solvers require vectors of the low-order-refined mesh coordinates, from which \textit{hypre} can build interpolation matrices that are used in the auxiliary space preconditioning method.
Additionally, AMS requires a \textit{discrete gradient matrix}, that maps functions in the low-order-refined $H^1$ finite element space (defined as values at the LOR mesh vertices) to their gradients in the corresponding $\Hcurl$ space (defined on edges of the LOR mesh).
Likewise, ADS requires a \textit{discrete curl matrix}, that maps functions in the low-order-refined $\Hcurl$ space to their curls in the corresponding $\Hdiv$ space (defined on faces of the LOR mesh).
Because of the construction of the interpolation--histopolation basis as described in \Cref{sec:vector-fe}, the high-order and low-order-refined versions of these matrices exactly coincide, and can be constructed using purely topological information.
That is, the matrices depend only on the topology of the high-order (coarse) mesh and the polynomial degree, and do not depend on the mesh geometry or other problem-specific parameters.

The LOR mesh vertex coordinate vectors can be computed efficiently using geometric information that is already available from the LOR matrix assembly in Step \ref{item:lor-assembly}.
The matrix assembly required the mesh coordinates in the form of a high-order E-vector;
this vector contains the mesh coordinates required for the construction of the interpolation matrices, with duplications because of the E-vector format.
Deduplicating this vector is performed efficiently using the element restriction degree of freedom index mappings (similarly to the action of $G^T$ in Step \ref{item:pa-apply});
this operation can be performed in parallel with one thread per deduplicated DOF without requiring any MPI communication.

The discrete gradient matrix maps from LOR vertices to edges, such that the value 1 or -1 is placed at the $(i,j)$ entry of the matrix, where the LOR vertex $j$ is one of the endpoints of the edge $i$, and the sign is determined by the orientation of the edge.
Because each LOR macro element has a uniform Cartesian structure, a local vertex-to-edge mapping can be computed once, and reused for the entire mesh.
The values are inserted into the correct locations using the index mappings from the element restriction operator.
Note that grad-div problems in 2D can also use the AMS solver, but with a modified gradient matrix that computes the \textit{rotated gradient} $\nabla^\perp = (-\partial_y, \partial_x)$.
The construction of this operator largely follows the same structure.
Pseudocode for the kernel used to construct the discrete gradient matrix in CSR format is shown in \Cref{alg:discrete-gradient}.
In this algorithm, the DOF and maps \texttt{element\_map}, \texttt{hcurl\_global\_to\_local} and \texttt{h1\_local\_to\_global} encode the information used to compute the action of the high-order $G$ operator (see \Cref{sec:pa-setup}).
In the low-order-refined context, these are interpreted as macro-element maps.
Finally, \texttt{edge\_to\_vertex} gives the indices of the two vertices corresponding to an (oriented) low-order-refined edge.
Because each macro-element has identical Cartesian structure, this one mapping is used for all DOFs.

\begin{algorithm}
   \caption{Construction of discrete gradient matrix in CSR format}
   \label{alg:discrete-gradient}
   \algrenewcommand\algorithmicfor{\textbf{parallel for}}
   \algrenewtext{EndFor}{\textbf{end}}
   \begin{algorithmic}[0]
      \State $N \gets \text{\# of \Nedelec DOFs}$
      \For{$i \in \{0, \ldots, N-1\}$}
         \State $e \gets \texttt{element\_map}[i]$
         \State $i_{\textit{loc}} \gets \texttt{hcurl\_global\_to\_local}[i]$
         \State $\sigma \gets \texttt{orientation}[i]$
         \State $j_{0,\textit{loc}} \gets \texttt{edge\_to\_vertex}[0,i_{\textit{loc}}]$
         \State $j_{1,\textit{loc}} \gets \texttt{edge\_to\_vertex}[1,i_{\textit{loc}}]$
         \State $I[i] \gets 2i$
         \State $J[2i] \gets \texttt{h1\_local\_to\_global}[e, j_{0,\textit{loc}}]$
         \State $J[2i+1] \gets \texttt{h1\_local\_to\_global}[e, j_{1,\textit{loc}}]$
         \State $A[2i] \gets -\sigma$
         \State $A[2i+1] \gets \sigma$
      \EndFor
      \State $I[N] \gets 2N$
   \end{algorithmic}
\end{algorithm}

Similarly, the discrete curl matrix maps from LOR edges to faces, such that the value 1 or -1 is placed at the $(i,j)$ entry of the matrix, where the LOR edge $j$ is one of the edges of the LOR face $i$.
The sign is determined by the orientation of the mesh edge relative to that of the mesh face.
The same consideration as in the case of the discrete gradient matrix also apply here;
a local version of the mapping is computed on a reference macro element, and reused for the entire mesh.
The index mapping information from the element restriction operator is used to assemble the global

\subsection*{Step \ref{item:amg-setup}. Algebraic multigrid setup}

After Step \ref{item:lor-assembly} has been performed, the parallel CSR matrix can be passed to algebraic multigrid library to perform the AMG setup.
In this work, we use \textit{hypre}'s AMG implementation, but any GPU-capable algebraic multigrid library could equally well be used, e.g.\ AmgX.
Because of the sparsity of the low-order-refined matrix, the construction of the AMG hierarchy is generally efficient, and the resulting operator complexities are not prohibitively high.
One of the main components of the algebraic multigrid setup is the construction of the coarse-grid operators, which is performed using triple-product ($P^T A P$) operations.
This is the same operation that is used for the global (parallel) LOR matrix assembly in Step \ref{item:asm-global}, and so both the algebraic multigrid setup and low-order-refined assembly benefit from improvements to the triple product algorithm.
Details on the development and porting of the AMG setup algorithms to GPU-based architectures are described in \cite{Falgout2021}.

\subsection*{Step \ref{item:pa-apply}. High-order operator evaluation}
\steplabel{pa-apply}

At each conjugate gradient iteration, the action of the high-order operator must be applied.
As described in Step \ref{item:pa-setup}, we use the \textit{partial assembly} approach for the high-order operator.
Given the decomposition \eqref{eq:op-decomp}, the action of the high-order operator $A_p$ can be computed by successively applying the operators $P$, $G$, $B$, $D$, and the transposes, $B^T$, $G^T$, $P^T$.
In the case of a conforming mesh, the matrices $P$ and $G$ are boolean matrices, which are stored as mappings between local and distributed degree of freedom indices; the case of nonconforming adaptive refinement results in more complicated prolongation operators that are treated differently.
The action of $P$ is computed using with MPI communication using these index mappings.
When possible, device-aware MPI is used for these communication routines in order to reduce memory transfer between host and device.
In MFEM, this is achieved using the \texttt{ConformingProlongationOperator} class.
The element restriction operator $G$, represented in MFEM with the class \texttt{ElementRestriction}, maps from rank-local vectors (referred to as L-vectors in CEED terminology) to element-wise vectors (referred to as E-vectors).
This is a type of gather operation that can be performed on device using a gather map, threaded over the number of E-vector degrees of freedom.
The action of the transpose $G^T$ is a scatter-type operation, that similarly can be performed efficiently on-device using the same threading strategy.

The action of the triple product $B^T D B$ is implemented as a single kernel that operates on the E-vector level.
In MFEM, this action is performed by an operator-specific class derived from \texttt{BilinearFormIntegrator}.
The specific form of the basis operator $B$ depends on the finite element bilinear form being evaluated.
For example, the mass matrix requires the evaluation of basis functions at quadrature points, for which $B$ takes the form of a Kronecker product,
\begin{equation} \label{eq:tensor-basis}
   B = \Bd \otimes \Bd \ \text{in 2D}, \qquad
   B = \Bd \otimes \Bd \otimes \Bd \ \text{in 3D},
\end{equation}
where $\Bd$ is the one-dimensional basis evaluation operator.
On the other hand, diffusion-type operators require the computation of the gradient of basis functions at quadrature points, for which $B$ can be written as
\[
   B = \begin{pmatrix}
      \Bd \otimes \Dd \\
      \Dd \otimes \Bd
   \end{pmatrix}\ \text{in 2D}, \qquad
   B = \begin{pmatrix}
      \Bd \otimes \Bd \otimes \Dd \\
      \Bd \otimes \Dd \otimes \Bd \\
      \Dd \otimes \Bd \otimes \Bd \\
   \end{pmatrix}\ \text{in 3D},
\]
where $\Dd$ is the one-dimensional basis differentiation operator.
Owing to the tensor-product structure of the $B$ operator, the action of $B$ and $B^T$ can be computed efficiently using sum factorization algorithms \cite{Orszag1980,Van-Loan2000}.
Only the one-dimensional matrix $\Bd$ and $\Dd$ are formed explicitly, and the action of the Kronecker product matrix-vector products is computed on the fly.

The most common threading strategy for the products $B^T D B$ is to use one block of threads per element, with threads corresponding to the quadrature points.
The basis function values (or gradients) at quadrature points are stored in shared memory, together with the intermediate vectors required for the sum factorized computation of the matrix-vector products.
The diagonal (or block diagonal) matrix $D$ is precomputed in Step \ref{item:pa-setup}, and its application trivially parallelizes over all quadrature points.
The transposed basis operator $B^T$ returns from quadrature points to (dual) degrees of freedom, and its application essentially follows to reverse order as that of $B$.

\subsection*{Step \ref{item:amg-solve}. Algebraic multigrid V-cycle}

The application of the algebraic multigrid V-cycle (i.e.\ the AMG solve phase) is more amenable to straightforward GPU implementation than the setup phase.
Each level in the multigrid hierarchy involves residual computation, point smoothing, and the application of restriction and prolongation matrices.
These operations can entirely be cast as standard linear algebraic operations, i.e.\ sparse matrix-vector (\texttt{spmv}) products and vector operations such as \texttt{axpy}.
In \textit{hypre}'s GPU AMG implementation, these sparse matrix-vector products are computed using vendor-provided libraries such as cuSPARSE and rocSPARSE.
At the coarsest level, the system is small enough that it can be efficiently solved directly.
For more details on the GPU acceleration of AMG methods, see \cite{Haase2010,Naumov2015,Falgout2021}.

\section{Results and numerical experiments}
\label{sec:results}

In this section, we present several numerical results studying the performance of the GPU algorithms described above.
These algorithms are implemented in the MFEM finite element software library \cite{Anderson2020}, see \Cref{sec:software}.
The low-order-refined discretizations are formed using the class \texttt{LORDiscretization} (and its parallel counterpart \texttt{ParLORDiscretization}), and the GPU-accelerated matrix assembly is performed by the class \texttt{BatchedLORAssembly}.
The class template \texttt{LORSolver} is used to create preconditioners using the LOR discretizations.
The specializations \texttt{LORSolver<HypreAMS>} and \texttt{LORSolver<HypreADS>} handle the construction of the coordinate vectors, and discrete gradient and discrete curl matrices required by the AMS and ADS solvers.
The algebraic multigrid solvers are provided by the \textit{hypre} library.
All results were performed on LLNL's \textit{Lassen} supercomputer, each node of which has 4 NVIDIA V100 GPUs and 44 Power 9 CPU cores.

\subsection{Algorithmic components}
\label{sec:results-components}

\begin{figure}
   \begin{tabular}{ll}
      \includegraphics[scale=0.9]{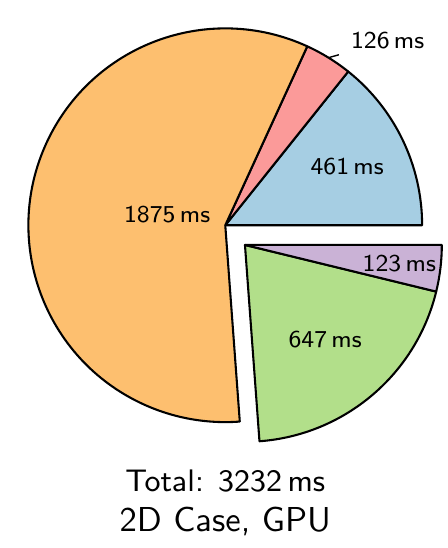} &
      \includegraphics[scale=0.9]{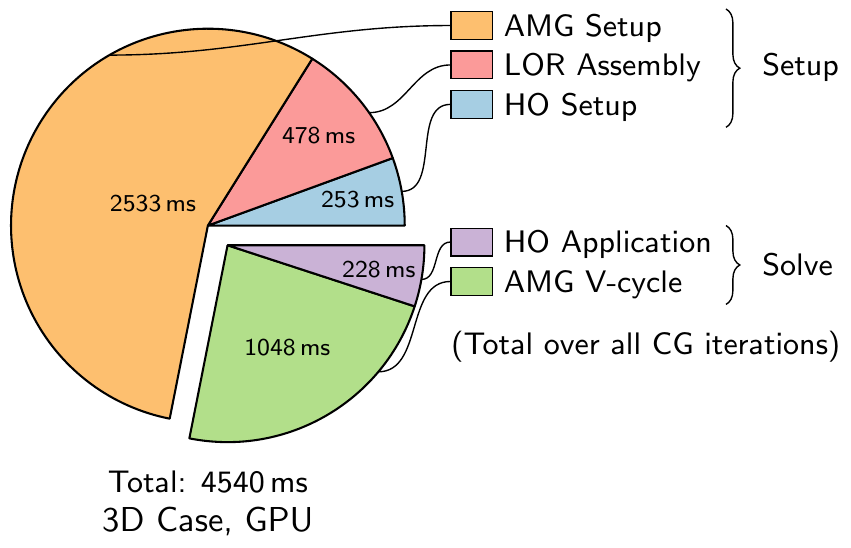} \\
      \includegraphics[scale=0.9]{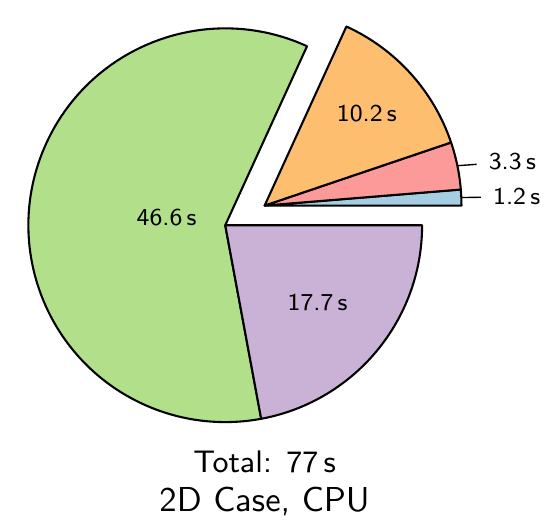} &
      \includegraphics[scale=0.9]{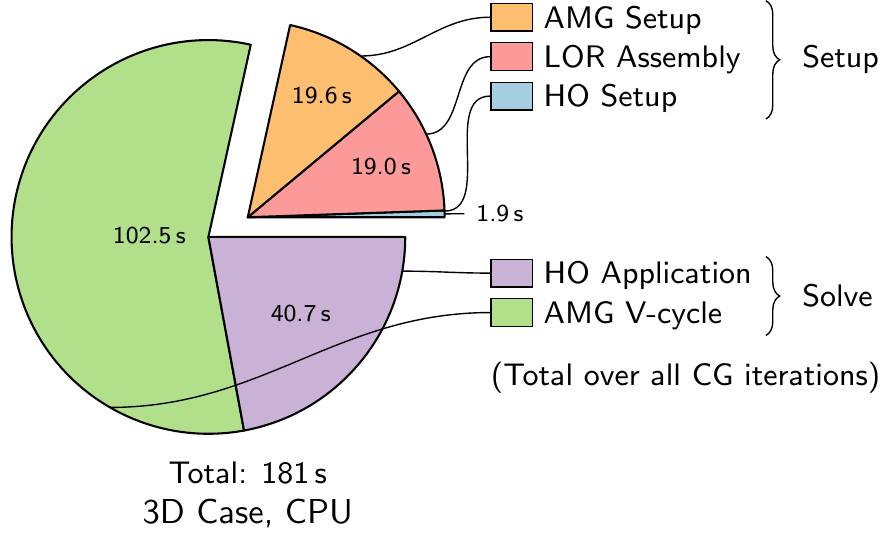}
   \end{tabular}

   \caption{
      Wall-clock runtimes for the algorithmic components described in \Cref{sec:algorithms}, solving a definite Helmholtz problem in $H^1$ with polynomial degree $p=6$.
      Top row: GPU timings on one V100 GPU.
      Bottom row: CPU timings on one Power 9 core.
      Left column: 2D test case with 262{,}144 elements and 9{,}443{,}329 degrees of freedom, 33 CG iterations.
      Right column: 3D test case with 32{,}768 elements at 7{,}189{,}057 degrees of freedom, 40 CG iterations.
   }
   \label{fig:pie}
\end{figure}

First we consider the relative weighting of the algorithmic components described in \Cref{sec:algorithms}.
We solve a definite Helmholtz (i.e.\ diffusion--reaction) problem on a Cartesian grid in 2D and 3D using polynomial degree $p=6$.
In 2D, the mesh consists of 262{,}144 elements, resulting in a total of 9{,}443{,}329 degrees of freedom.
In 3D, the mesh consists of 32{,}768 elements, resulting in 7{,}189{,}057 degrees of freedom.
Because in both cases the mesh is topologically Cartesian, the low-order-refined system matrix has 9 nonzeros per row in 2D, and 27 nonzeros per row in 3D (with the exception of degrees of freedom at the domain boundary), resulting in 84{,}953{,}089 nonzeros in 2D and 192{,}100{,}033 nonzeros in 3D.
The linear system is solved using a relative tolerance of $10^{-12}$.
In 2D, this requires 32 CG iterations, and in 3D convergence is attained after 40 CG iterations.
These problems are solved on a single V100 GPU.
We also compare the GPU timings with the runtimes on one Power 9 CPU core.
We emphasize that the purpose of this comparison is to study the difference in \textit{relative weightings} of the algorithmic components on CPU and GPU, not to compare directly GPU to CPU performance.
The runtimes for the algorithmic components of the solution procedure are shown in \Cref{fig:pie}.

On the GPU, in both the 2D and 3D cases, around 75\% of the time to solution was spent in the setup phase, with over 50\% of the time to solution in AMG setup.
The time spent assembling the LOR system was approximately 3\% and 10\% of the total runtime, in 2D and 3D respectively.
Note that the timings for the setup of the high-order operator also includes the construction of the element restriction operator (see \Cref{sec:pa-setup}), which, because of the macro-element strategy described in \Cref{sec:lor-assembly}, is reused for the LOR matrix assembly.
Within the solve phase, over 80\% of the runtime was spent in the AMG V-cycle in both cases, and only 16--18\% in the high-order operator evaluation.
These results suggest that the largest potential for further performance gains are possible by optimizing the preconditioner construction and application.
Interestingly, on CPU, the relative weight of the AMG setup is significantly reduced, and in contrast to the GPU runs, the large majority of the time is spent in the solve phase.

These results highlight one advantage of LOR-based methods: any improvements or optimizations to the algebraic multigrid implementation (or in fact any other preconditioner suitable for the low-order problem) can automatically benefit the high-order solvers.
We further note that in many practical time-dependent problems, the mesh and problem coefficients remain fixed, either for the duration of the entire simulation, or for numerous time steps, see, e.g.\ \cite{Franco2020}.
In this context, the setup time is amortized over the many repeated applications of the solve phase.
As a result, for these problems, the relatively expensive AMG setup may constitute a small fraction of the overall runtime.
However, for problems which require performing the preconditioner setup every time step (e.g.\ problems incorporating mesh motion or time-dependent variable coefficients), the setup will remain a significant portion of the time to solution.

\subsection{Kernel throughput}

\begin{figure}
   \includegraphics[width=0.33\linewidth]{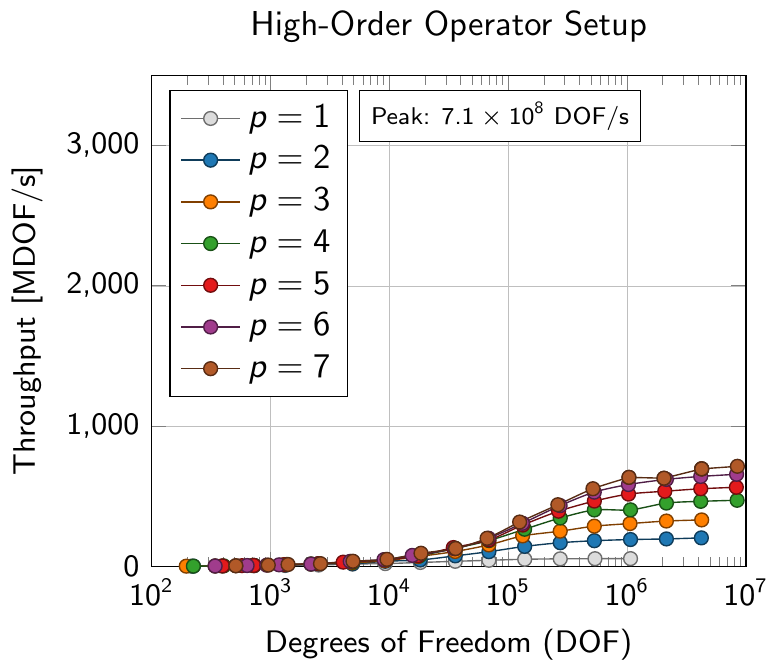}%
   \includegraphics[width=0.33\linewidth]{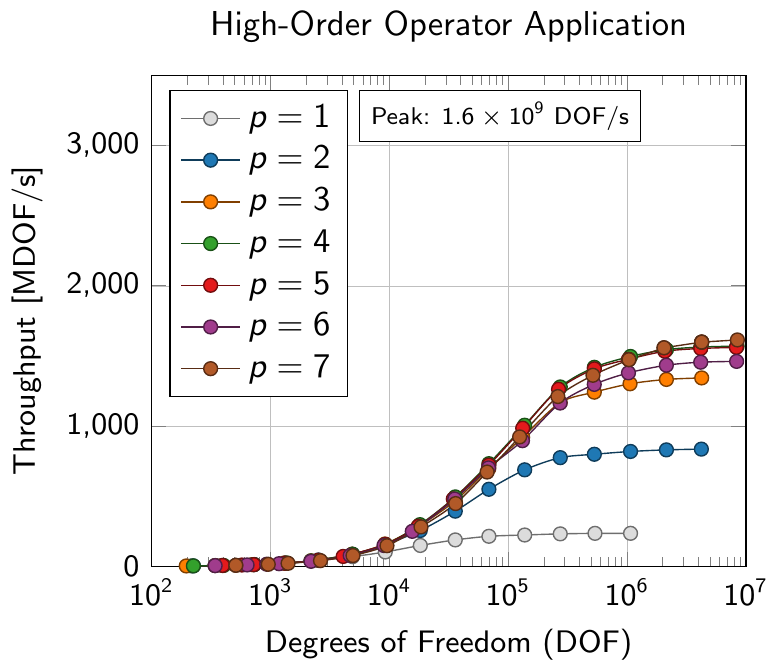}%
   \includegraphics[width=0.33\linewidth]{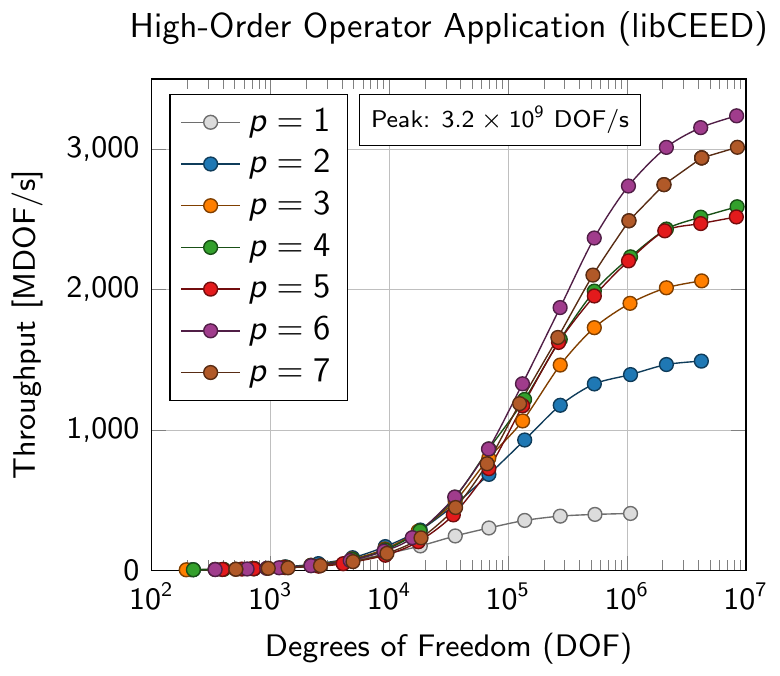}

   \caption{
      Throughput (millions of degrees of freedom per second) for high-order operator and setup for the definite Helmholtz problem on one V100 GPU.
      Left: high-order operator setup (partial assembly).
      Center: high-order operator evaluation using MFEM's native kernels.
      Right: high-order operator evaluation using MFEM's libCEED backend.
   }
   \label{fig:throughput-ho}
\end{figure}

We study the dependence of the kernel throughput on polynomial degree and problem size.
Working on a sequence of 3D Cartesian meshes, we measure the throughput (degrees of freedom per second) of the main computational kernels.
First, we consider high-order operator setup and application.
The MFEM library supports multiple backends that can be used for GPU evaluation.
We compare MFEM's native GPU backend to libCEED's \texttt{cuda-gen} backend, which uses code generation to fuse the processor-local products $G^T B^T D B G$ (cf.\ \Cref{eq:op-decomp} and reference \cite{Brown2021}).
The results are shown in \Cref{fig:throughput-ho}.
These high-order operations are independent of the choice of preconditioner, and have previously been studied in detail, see \cite{Kolev2021,Kolev2021a,Fischer2020,Franco2020}.
As is typical with high-order finite elements, greater throughput is achieved for higher polynomial degrees.
For $p=7$, the operator setup (i.e.\ the construction of the $D$ matrix described in \Cref{sec:pa-setup}) reaches a peak throughput for about 700 million degrees of freedom per son, and MFEM's native operator application (using the algorithm described in \Cref{sec:pa-apply}) reaches a peak throughput of 1.6 billion degrees of freedom per second, while libCEED's operator evaluation reaches a peak of 3.2 billion degrees of freedom per second.

\begin{figure}
   \includegraphics[width=0.33\linewidth]{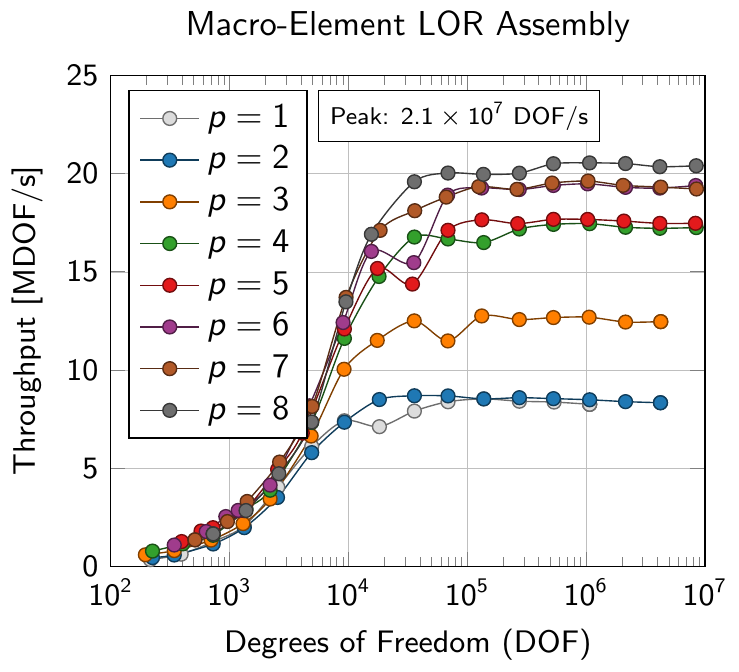}%
   \includegraphics[width=0.33\linewidth]{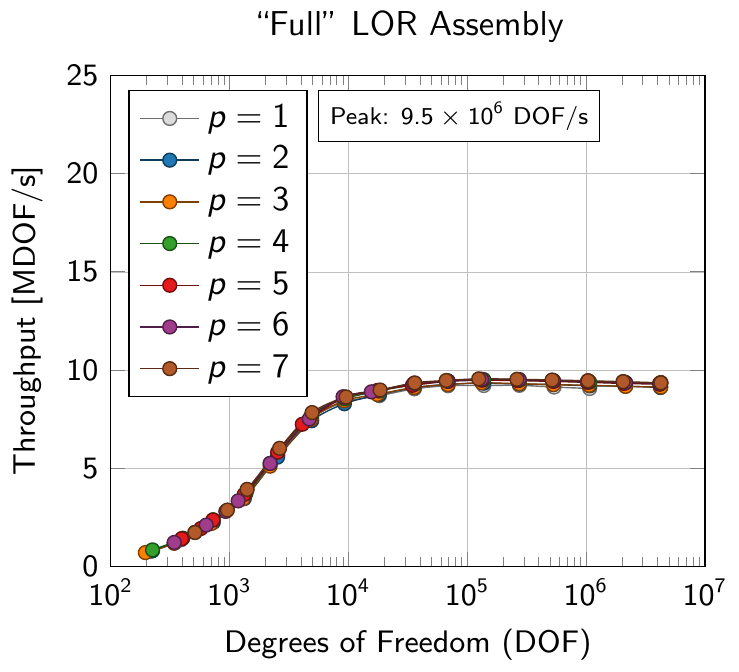}%
   \includegraphics[width=0.33\linewidth]{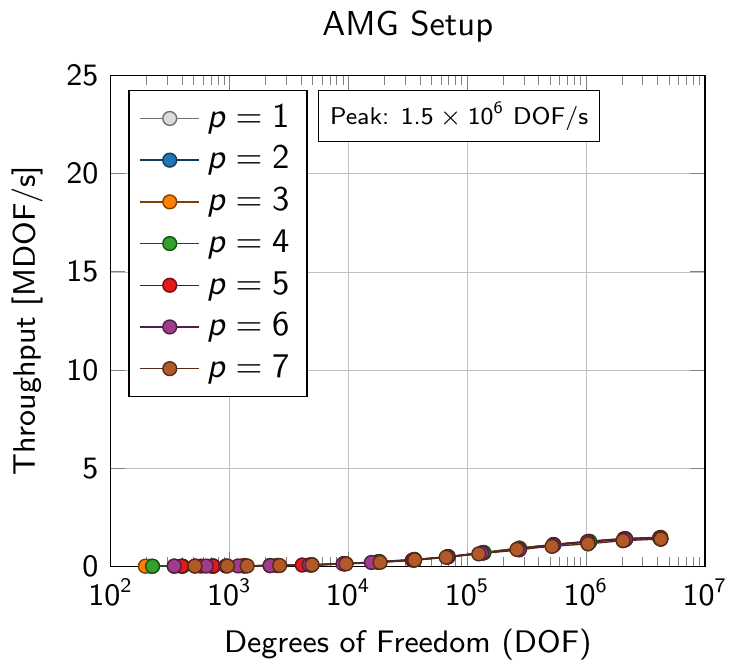}

   \caption{
      Throughput (millions of degrees of freedom per second) for high-order operator and setup for the definite Helmholtz problem on one V100 GPU.
      Left: LOR matrix assembly using the macro-element approach.
      Center: LOR matrix assembly using the unstructured (legacy) approach.
      Right: algebraic multigrid setup.
   }
   \label{fig:throughput-assembly-amg}
\end{figure}

Next, we consider the throughput of the low-order-refined assembly algorithms.
We compare the macro-element assembly strategy described in \Cref{sec:lor-assembly} to a more generic, fully unstructured matrix assembly algorithm that does not take advantage of the structure of the low-order-refined mesh.
The more general, unstructured algorithm is suitable for assembling matrices corresponding to any polynomial degree, but does not contain any optimizations specific to the case of low-order-refined discretizations.
This algorithm uses one block of threads per element, with one thread per degree of freedom, constructing small dense matrices for each element of the low-order-refined mesh, and then directly assembling the global sparse matrix (skipping the macro-element sparse matrix assembly in the LOR approach).
The throughput plots comparing these two approaches are shown in the left and center columns of \Cref{fig:throughput-assembly-amg}.

The fully unstructured assembly algorithm shows no variation in throughput between polynomial degrees of the high-order space;
this is expected because for equal problem size, the low-order-refined meshes corresponding different high-order polynomial degrees are topologically identically, and differ only in the mesh distortion, which has no algorithmic impact on the matrix assembly.
On the other hand, the macro-element assembly algorithm exhibits increasing throughput with high-order polynomial degree.
As the degree of the high-order space increases, the size of each macro-element increases, and the matrix assembly becomes more structured, allowing for increased fine-grained parallelism, leading to higher performance.
In the lowest-order case, the macro element strategy performs worse than the unstructured algorithm.
This is because in the case of $p=1$, the macro-elements each only contain a single element, and so each block consists only of one thread.
On the other hand, the unstructured algorithm threads over the E-vector degrees of freedom, resulting in 8 threads per element.
However, at $p=2$, the performance of the macro-element strategy is roughly equal to that of the unstructured algorithm, and for $p > 2$ we observe meaningful performance improvements.
For the highest orders, the macro-element approach is more than twice as fast as the unstructured approach.
It is also important to note that this factor does not include the savings that are obtained by reusing the high-order mesh topology and element restriction operator.
In the fully unstructured case, the overhead associated with constructing the low-order-refined mesh and element restriction operators can dominate the time spent in the matrix assembly, often by large factors.

\begin{figure}
   \includegraphics[width=0.33\linewidth]{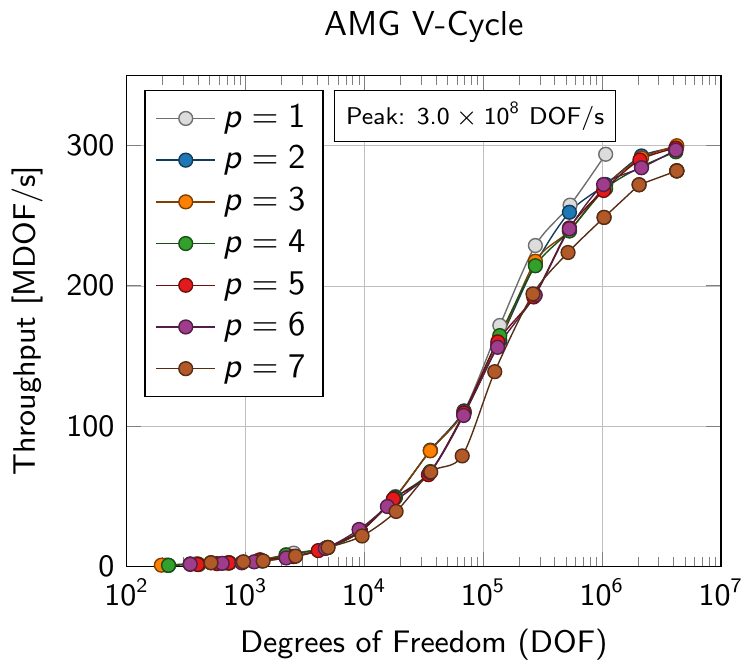}%
   \includegraphics[width=0.33\linewidth]{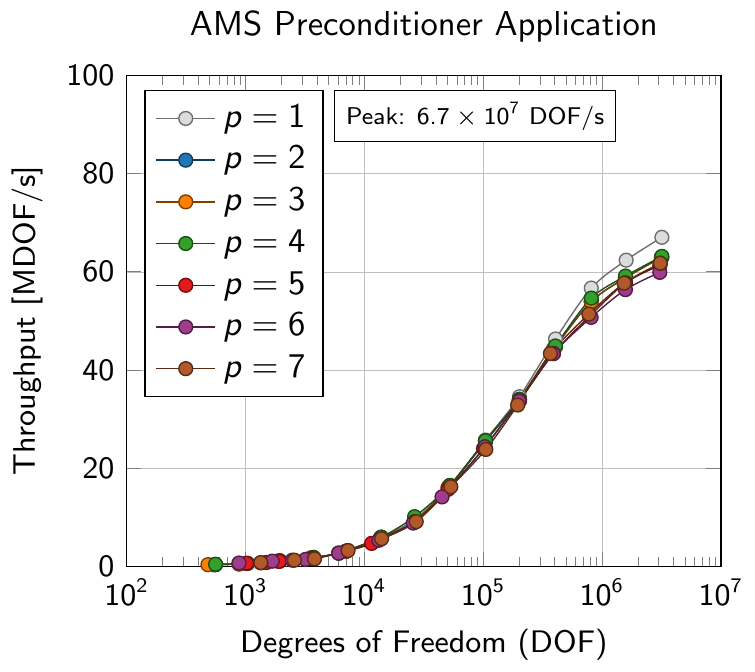}%
   \includegraphics[width=0.33\linewidth]{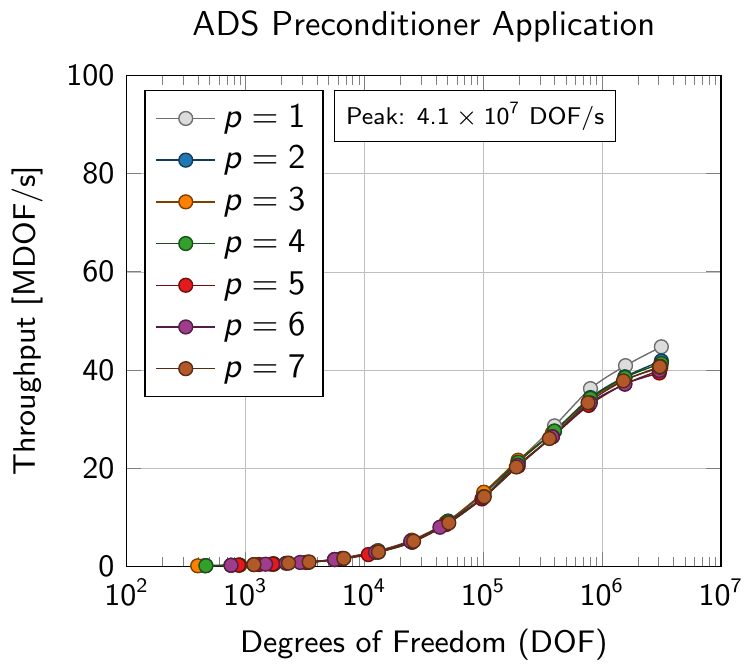}

   \caption{
      Throughput (millions of degrees of freedom per second) for algebraic multigrid preconditioner application on one V100.
      Left: one BoomerAMG V-cycle.
      Center: application of the AMS preconditioner for $\Hcurl$.
      Right: application of the ADS preconditioner for $\Hdiv$.
   }
   \label{fig:throughput-vcycle}
\end{figure}

The right column of \Cref{fig:throughput-assembly-amg} shows the throughput of the algebraic multigrid setup as a function of problem size.
As in the case of the unstructured LOR matrix assembly, there is essentially no dependence of the performance of these kernels on the polynomial degree of the high-order space.
The resulting system matrices have the same sparsity pattern, independent of $p$.
Differences in the values of the matrix entries can lead to different choices in the AMG coarsening, and hence will result in different AMG hierarchies, which has the potential to have a marginal impact on performance.
In practice, these differences are found to be negligible.
The results obtained here corroborate those shown in \Cref{sec:results-components}.
The bottleneck in the problem setup consists of the AMG setup phase, which reaches a peak throughput of about 1.5 million degrees of freedom for the largest problem sizes.
Finally we consider the AMG V-cycle, the throughput of which is shown in the left column of \Cref{fig:throughput-vcycle}.
As in the case of AMG setup, the performance of these operations does not display any dependence on the polynomial degree.
This is because the sparsity patterns for LOR systems of a given size are independent of the polynomial degree of the high-order space.
The BoomerAMG V-cycle reaches a peak throughput of about 300 million degrees of freedom per second.

\subsection{\Nedelec and Raviart--Thomas elements}

\begin{figure}
   \renewcommand{\tabcolsep}{0pt}
   \begin{tabular}{ccc}
      \includegraphics[scale=0.68]{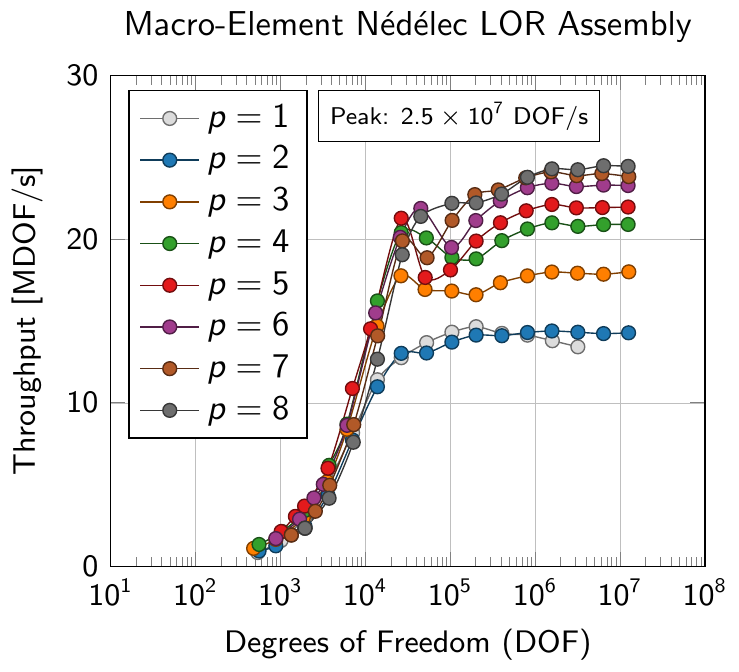}&%
      \includegraphics[scale=0.68]{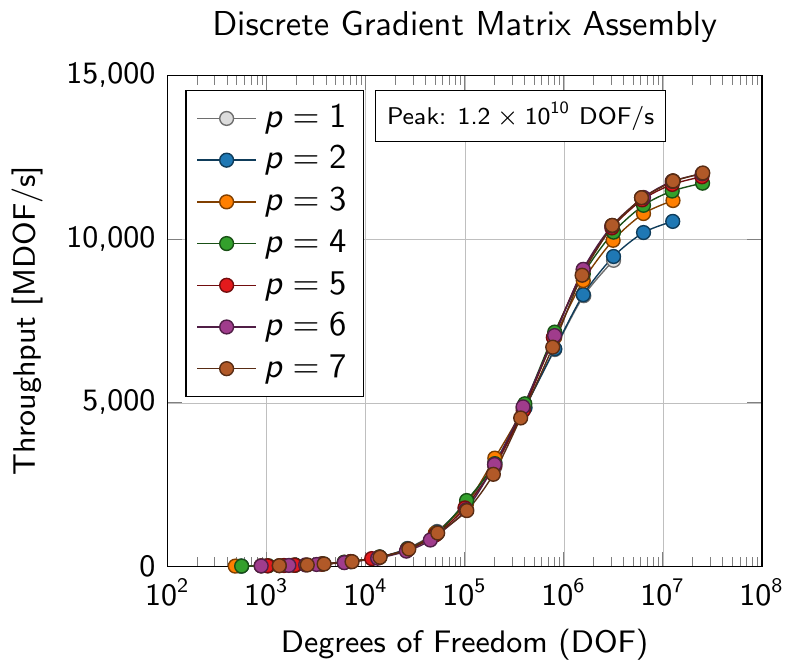}&%
      \includegraphics[scale=0.68]{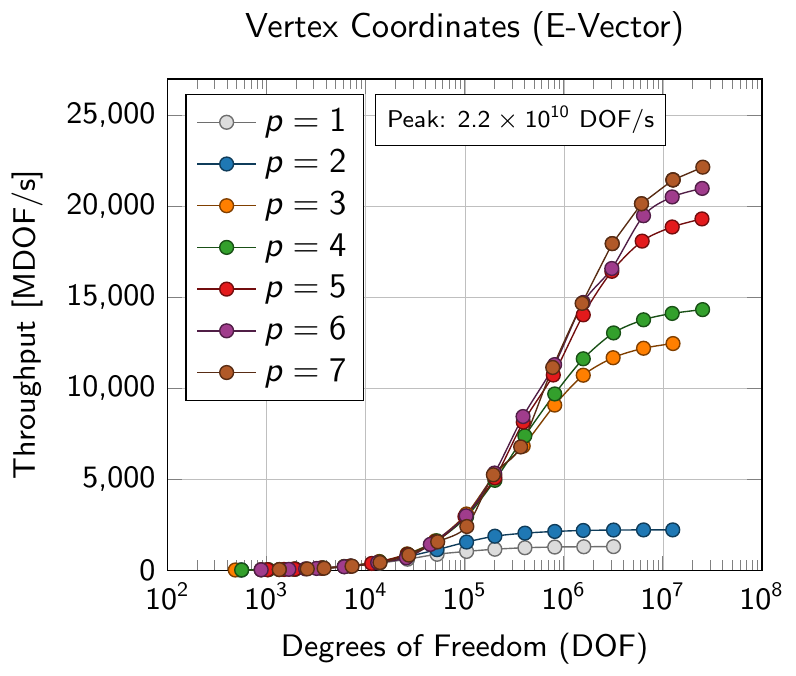}\\
      \includegraphics[scale=0.68]{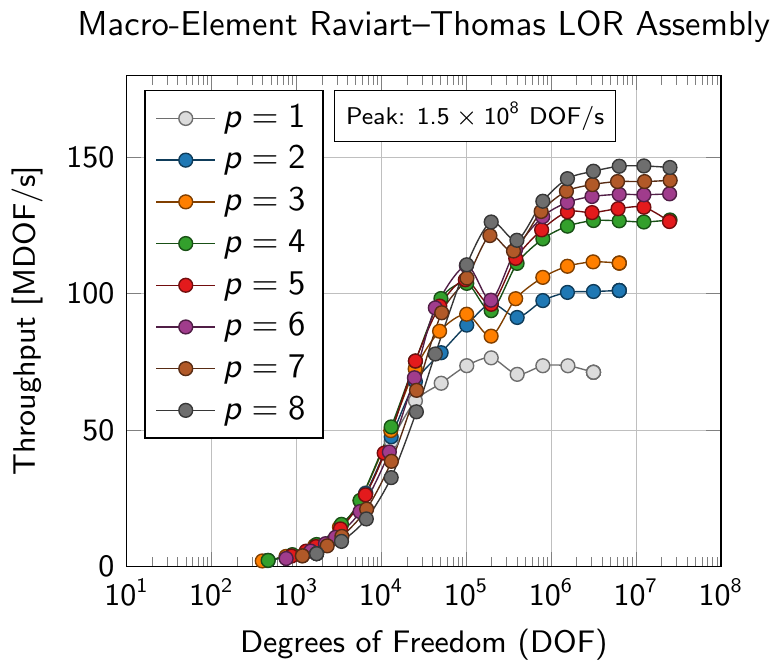}&%
      \includegraphics[scale=0.68]{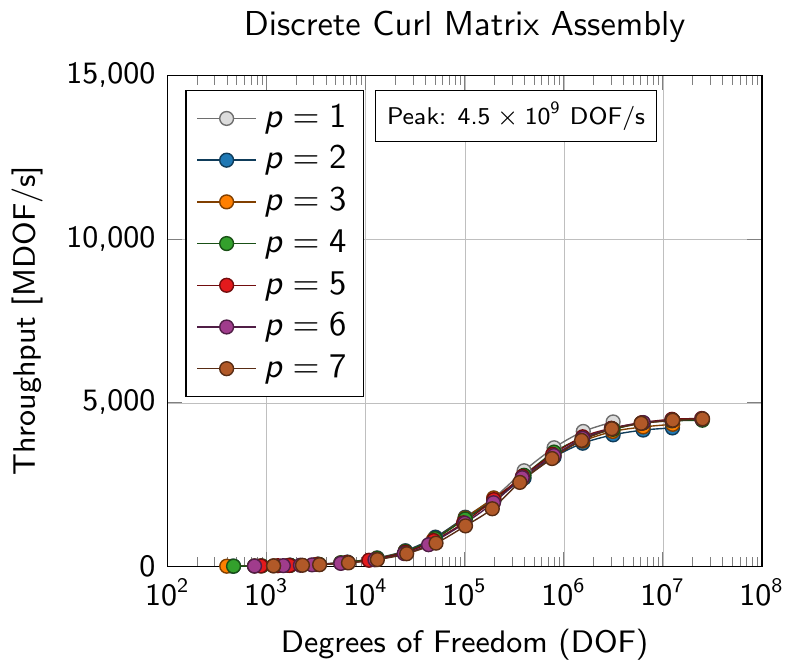}&%
      \includegraphics[scale=0.68]{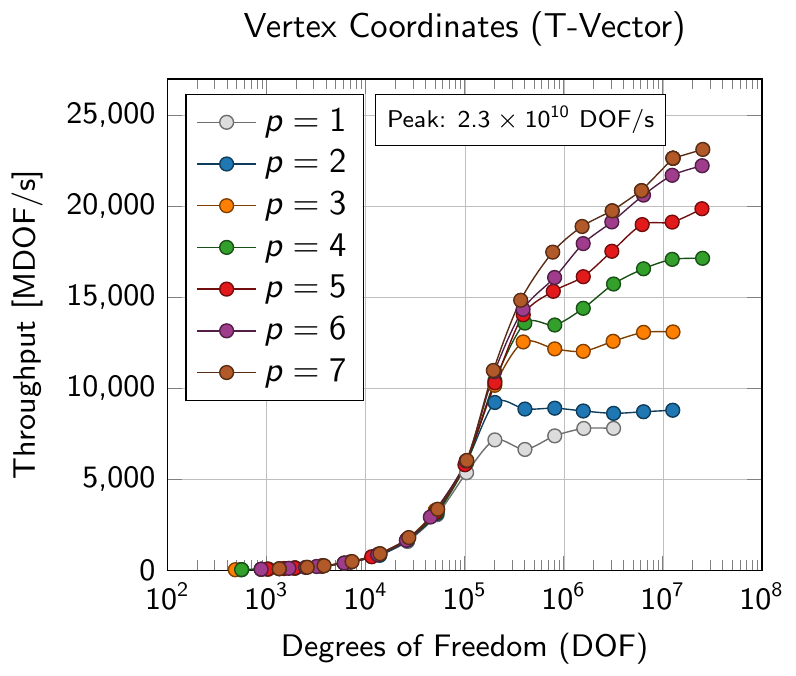}\\
   \end{tabular}
   \caption{
      Throughput (millions of degrees of freedom per second) for \Nedelec and Raviart--Thomas problems.
      Left column: assembly of LOR system matrix.
      Center column: construction of discrete gradient and curl matrices.
      Right column: construction of coordinate vectors needed for discrete interpolation.
   }
   \label{fig:throughput-nd-rt}
\end{figure}

In this section, we consider the throughput of the kernels required to construct low-order-refined preconditioners for problems in $\Hcurl$ and $\Hdiv$ using \Nedelec and Raviart--Thomas elements.
First, we consider the assembly of the LOR system matrix, for which throughput plots are shown in the left column of \Cref{fig:throughput-nd-rt}.
The assembly of the LOR \Nedelec matrices performs roughly the same as the $H^1$ assembly (compare with \Cref{fig:throughput-ho}).
However, the Raviart--Thomas assembly enjoys significantly higher throughput.
This is largely because each Raviart--Thomas degree of freedom corresponds to a face of the low-order-refined mesh, whereas each \Nedelec degree of freedom corresponds to an edge, and each $H^1$ degree of freedom corresponds to a vertex.
Therefore, each RT DOF is shared by at most two elements, whereas in the case of a Cartesian mesh, \Nedelec and $H^1$ DOFs are shared by 4 and 8 elements, respectively.
As a result, the RT system matrix enjoys greater sparsity, and the sparse matrix construction requires fewer atomic additions during assembly.

As described in \Cref{sec:vector-fe}, the AMS and ADS algebraic auxiliary space solvers for $\Hcurl$ and $\Hdiv$ problems require the construction of discrete interpolation, gradient, and curl matrices.
The interpolation matrix is constructed using vectors of the mesh vertex coordinates, and the discrete gradient and curl are constructed using the mesh topology as described in \Cref{sec:algorithms}.
The throughput for the construction of the discrete gradient and curl is shown in the center and right columns of \Cref{fig:throughput-nd-rt}.
The peak throughput for the construction of the discrete gradient matrix is about 12 billion degrees of freedom per second, and the peak throughput for the construction of the discrete curl matrix is 4.5 billion degrees of freedom per second.
To construct the coordinate vectors needed for the interpolation operator, first the high-order (coarse) mesh vertices are interpolated to the low-order-refined (Gauss--Lobatto) mesh vertices in E-vector format.
This operation is essentially an element-wise tensor contraction using the tensor-product basis operator given in \eqref{eq:tensor-basis}.
The throughput for this operation is shown in the top-right plot of \Cref{fig:throughput-nd-rt}.
Once the coordinate vector has been obtained in E-vector format, it should be converted into T-vector (global, parallel) format.
Conceptually, this operation corresponds to applying the element restriction operator to convert from E-vector to L-vector, and then applying the parallel restriction operator to convert from L-vector to T-vector.
In practice, this can be done in a single step without any parallel communication.
The throughput for this operation is shown in the bottom-right plot of \Cref{fig:throughput-nd-rt}.
Both coordinate vector operations achieve a peak throughput of over 22 billion degrees of freedom per second.
The throughput for the application of the AMS and ADS preconditioners is shown in the center and right columns of \Cref{fig:throughput-vcycle}.
The AMS and ADS preconditioners have lower throughput than one application of BoomerAMG, since each application of AMS and ADS requires multiple AMG V-cycles, in addition to applications of the discrete gradient, curl, and interpolation matrices (see \cite{Kolev2009,Kolev2012}).
The peak throughput of AMS is 67 million degrees of freedom per second, and of ADS is 41 million degrees of freedom per second.

\subsection{Parallel scaling}

In this section, we examine the parallel scalability of these algorithms.
The construction of the process-local sparse matrices is performed in independently on each MPI rank.
These matrices are then placed in a global block-diagonal matrix, with one block per rank.
The global system matrix is computed as $A = P^T A_L P$, where $A_L$ is the block diagonal matrix described above, and $P$ is the parallel prolongation matrix.
For this triple product, we use the sparse matrix-matrix product GPU implementation available in \textit{hypre}, the development of which is described in \cite{Falgout2021}.

Once the global (parallel) sparse matrix has been constructed, the remaining parallel operations are AMG setup, AMG V-cycle, and high-order operator evaluation.
For details on the parallelization of the BoomerAMG preconditioner, see \cite{Henson2002}.
The approach taken to parallelize the high-order operator is described in \Cref{sec:pa-apply}, using the operator decomposition \eqref{eq:op-decomp}.
Writing $A_p = P^T G^T B^T D B G P$, the process-local operator is given by $A_L = G^T B^T D B G$.
The action of this linear operator is completely local, and requires no MPI communication.
Therefore, the only operations that require parallel communication are the application of $P$ and its transpose.
For conforming meshes, the operator $P$ is represented in MFEM as an object of type \texttt{ConformingProlongationOperator}.
This class provides an optimized implementation of the action of $P$ and $P^T$ using device-aware MPI.
When the mesh is nonconforming, $P$ takes a more complicated form, and is represented as a parallel sparse matrix.
In this case, the action of $P$ and $P^T$ are computed as parallel sparse matrix-vector products using the \textit{hypre} library.

\begin{figure}
   \renewcommand{\tabcolsep}{0pt}
   \begin{tabular}{ccc}
      \includegraphics[scale=0.7]{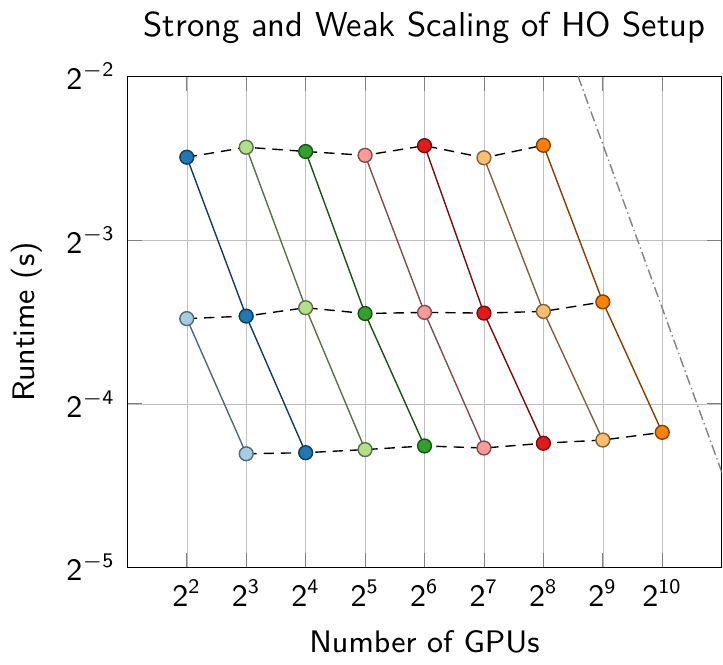}&
      \includegraphics[scale=0.7]{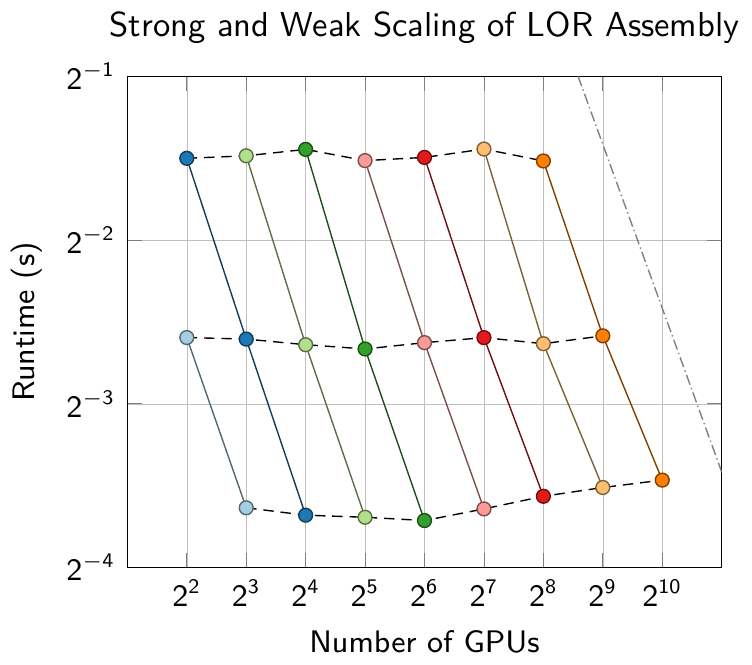}&
      \includegraphics[scale=0.7]{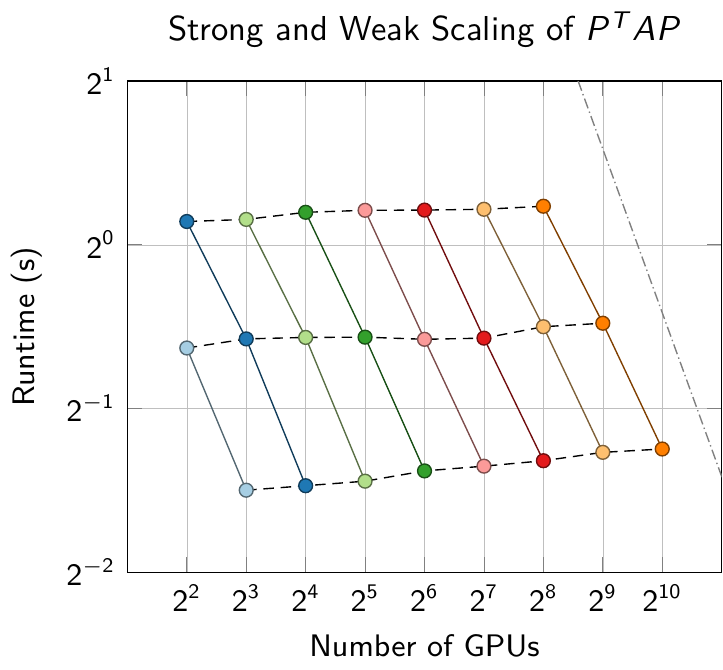}\\
      \includegraphics[scale=0.7]{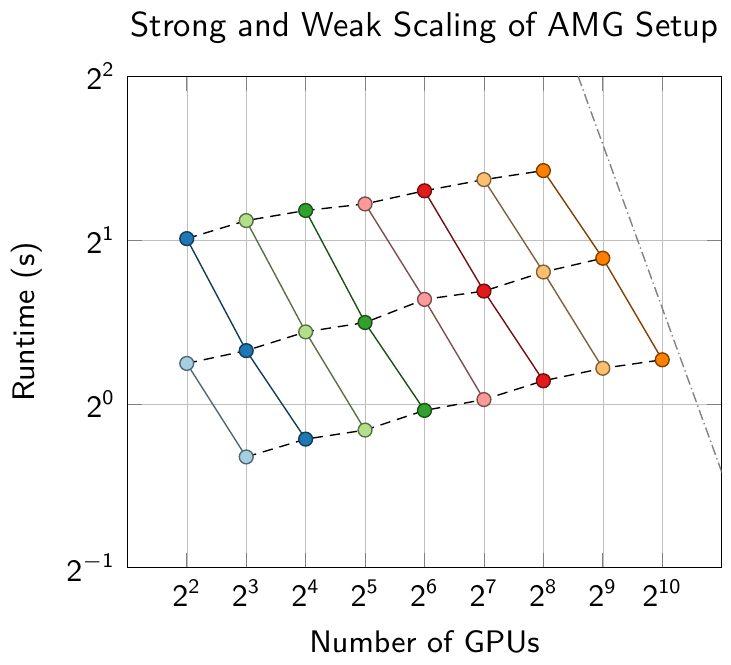}&
      \includegraphics[scale=0.7]{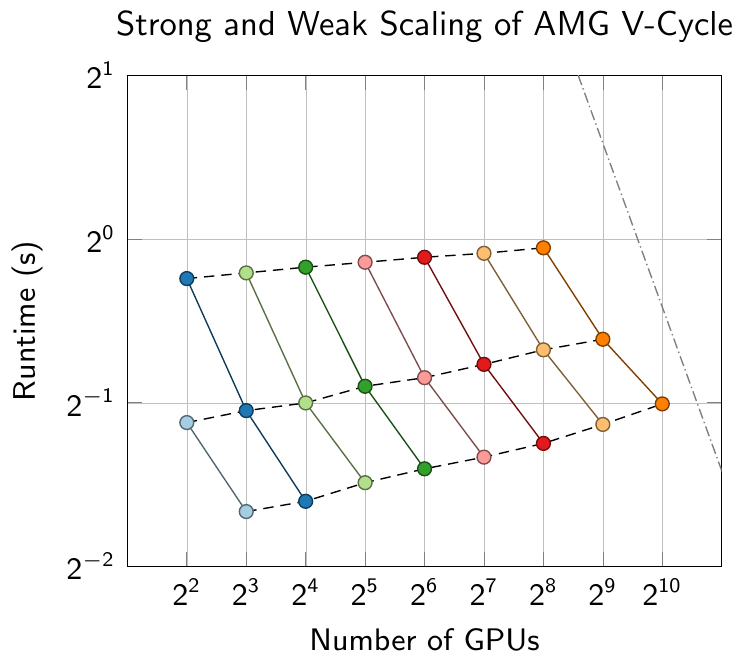}&
      \includegraphics[scale=0.7]{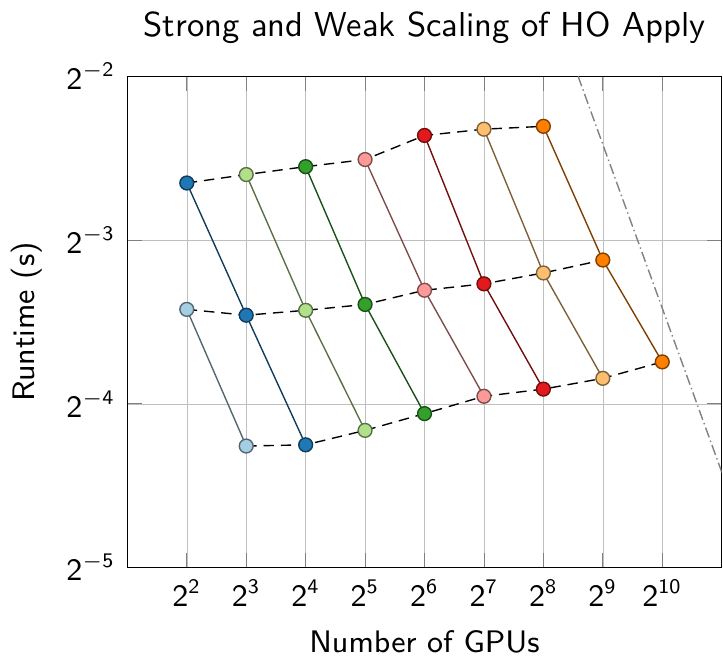}\\
      \multicolumn{3}{c}{%
         \raisebox{-0.5\height}{\includegraphics[scale=0.9]{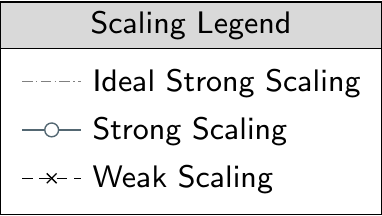}}%
         \hspace{0.25cm}%
         \raisebox{-0.5\height}{\includegraphics[scale=0.9]{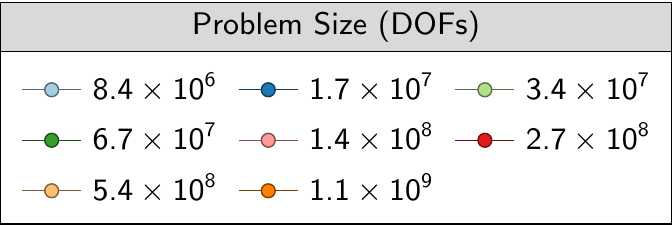}}%
      }
   \end{tabular}
   \caption{
      Parallel scaling results for setup and solve phases from 4 to 1024 Tesla V100-SXM2 NVIDIA GPUs.
      Strong scaling results are given by the solid lines, and weak scaling results are given by the dashed lines.
      The ideal strong scaling curve is shown for reference.
      Ideal weak scaling corresponds to horizontal lines.
      The results have been done with CUDA 11.7.0 and \textit{hypre} 2.25.0, with optimized sparse matrix-matrix multiplications (SpGEMM).
   }
   \label{fig:scaling}
\end{figure}

To measure the parallel scalability of our algorithms, we perform a strong and weak scaling study, using between 4 and 1024 GPUs.
We consider 8 problem configurations, with sizes increasing by factors of two from $8.4 \times 10^6$ degrees of freedom to $1.1 \times 10^9$ degrees of freedom.
Each problem configuration is run on three sets of GPUs to obtain the strong scaling results.
Collecting the timings across different problem configurations gives the weak scaling results.
The different algorithmic components as outlined in the previous sections are instrumented separately.
The scaling results are shown in \Cref{fig:scaling}.
Comparisons with ideal strong and weak scaling curves indicate excellent parallel scalability and efficiency for the high-order operator setup, low-order-refined matrix assembly, and high-order operator evaluation.
The algebraic multigrid setup is the component of the algorithm that is most challenging for parallel scalability.
The AMG V-cycle is less efficient for smaller problem sizes, but demonstrates good weak scalability for larger problem sizes.
The overall weak scaling parallel efficiency for the $1.1\times10^9$ problem on 256 GPUs was 83\% for the setup phase (HO set, LOR assembly, and AMG setup) and 87\% for the solve phase (HO apply and AMG V-cycle).

\subsection{Adaptive mesh refinement} \label{sec:amr-results}

\begin{figure}
   \includegraphics[height=5cm]{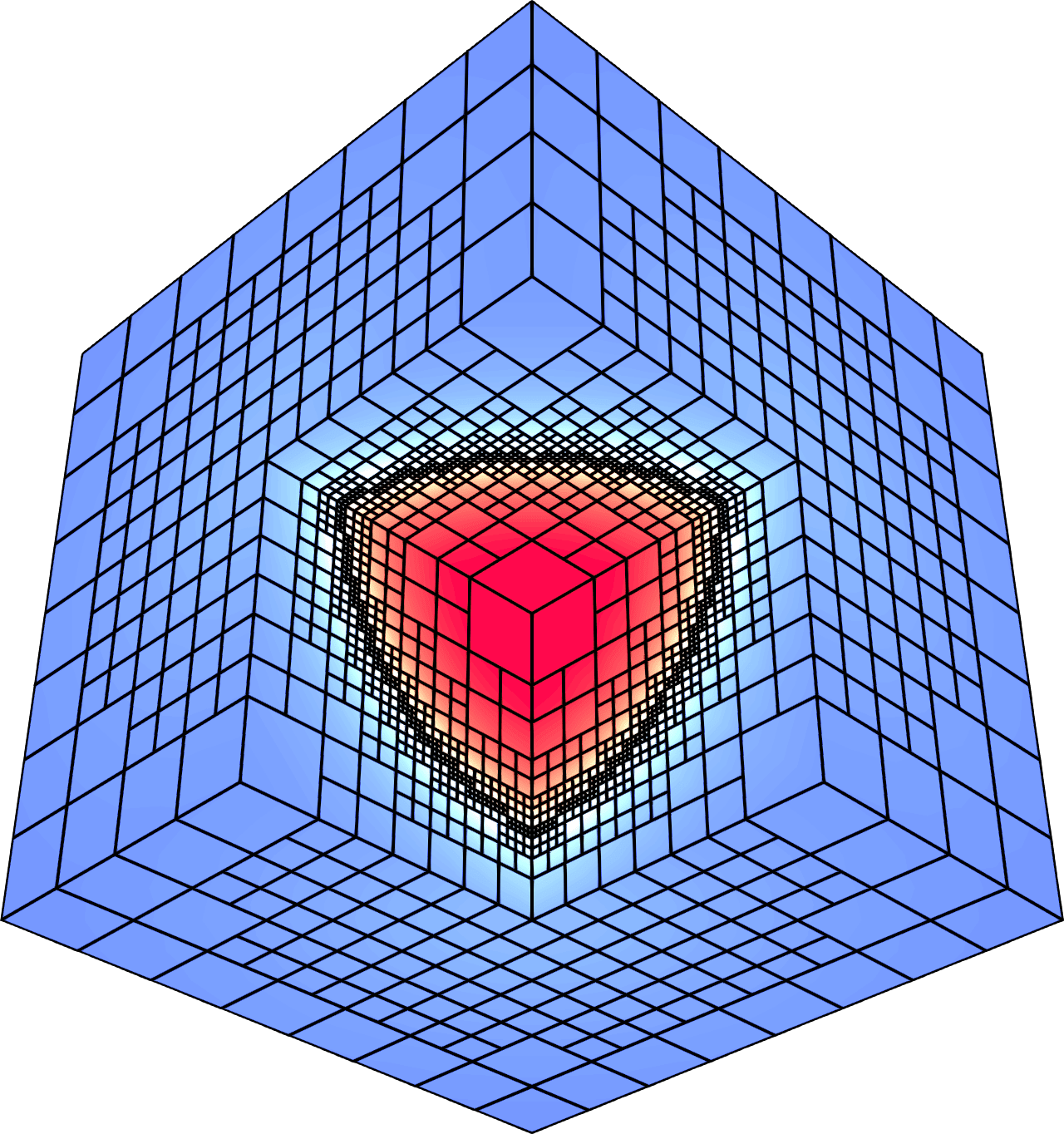}%
   \hspace{2cm}%
   \includegraphics[height=5cm]{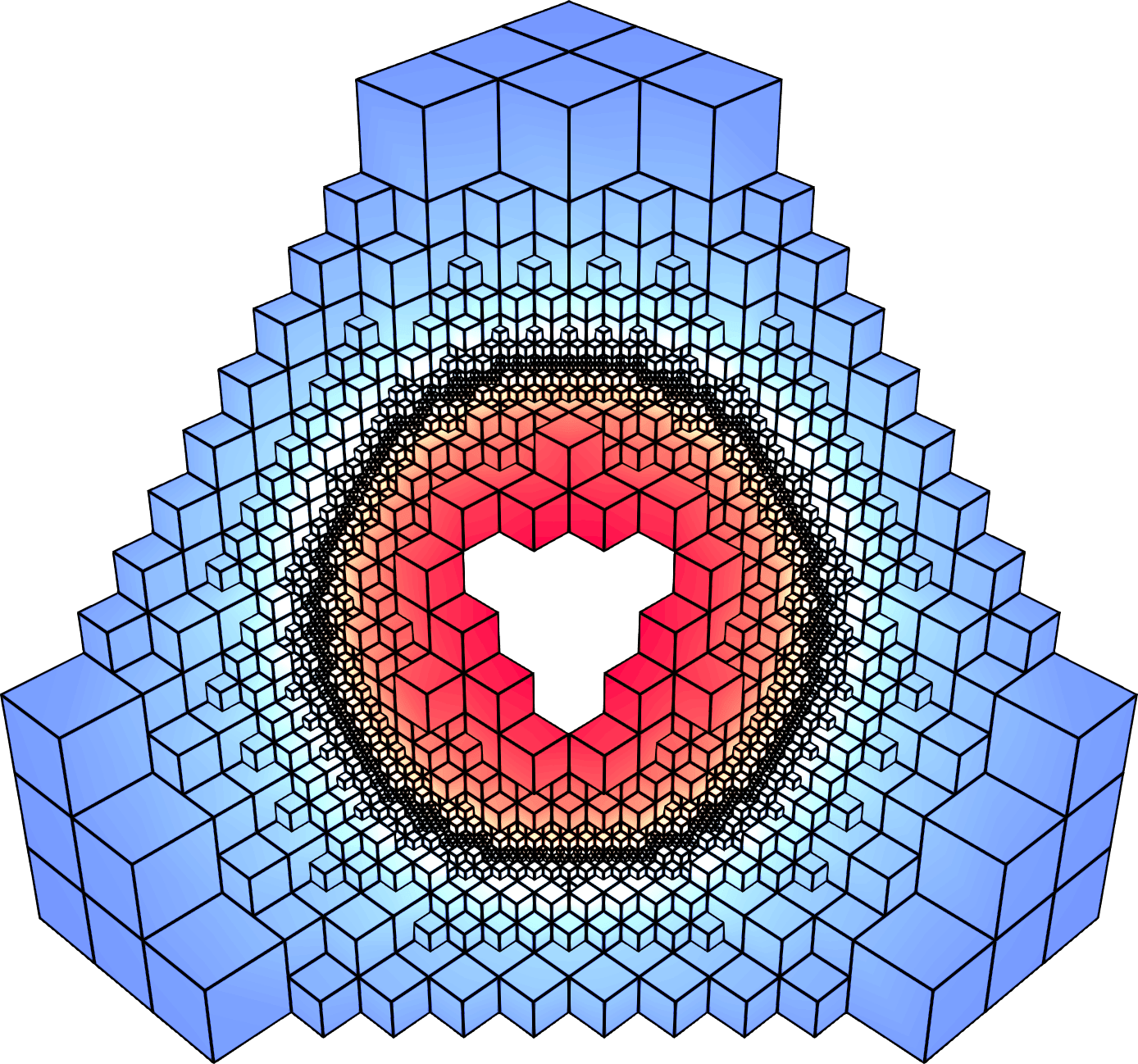}

   \caption{
      Mesh and solution of inner layer problem run, showing nonconforming adaptive mesh refinement, run on 12 V100 GPUs.
      Left: Fichera corner mesh.
      Right: mesh cut by plane normal to $(1,1,1)$ passing through the point $(0.2, 0.2, 0.2)$.}
   \label{fig:amr-mesh-solution}
\end{figure}

To illustrate the performance of the low-order-refined solvers on a problem with adaptive mesh refinement, we consider a Poisson problem whose solution exhibits an inner layer with a sharp gradient.
Let $u = \arctan(\alpha (r - r_0))$, where $r$ is the radial distance, and $\alpha$ is a sharpness parameter.
We choose $\alpha = 20$ and $r = 0.725$.
The right-hand side $f$ and Dirichlet condition $g_D$ are chosen such that $u$ is the solution to the Poisson problem \eqref{eq:poisson}.
We solve this problem on a mesh of the Fichera corner, which has been adaptively refined around the inner layer.
The refined mesh is 1-irregular, and consists of 93{,}940 elements with 23{,}562 nonconforming interfaces.
The solution and mesh are shown in \Cref{fig:amr-mesh-solution}.

\begin{figure}
   \small
   \raisebox{-0.5\height}{\includegraphics{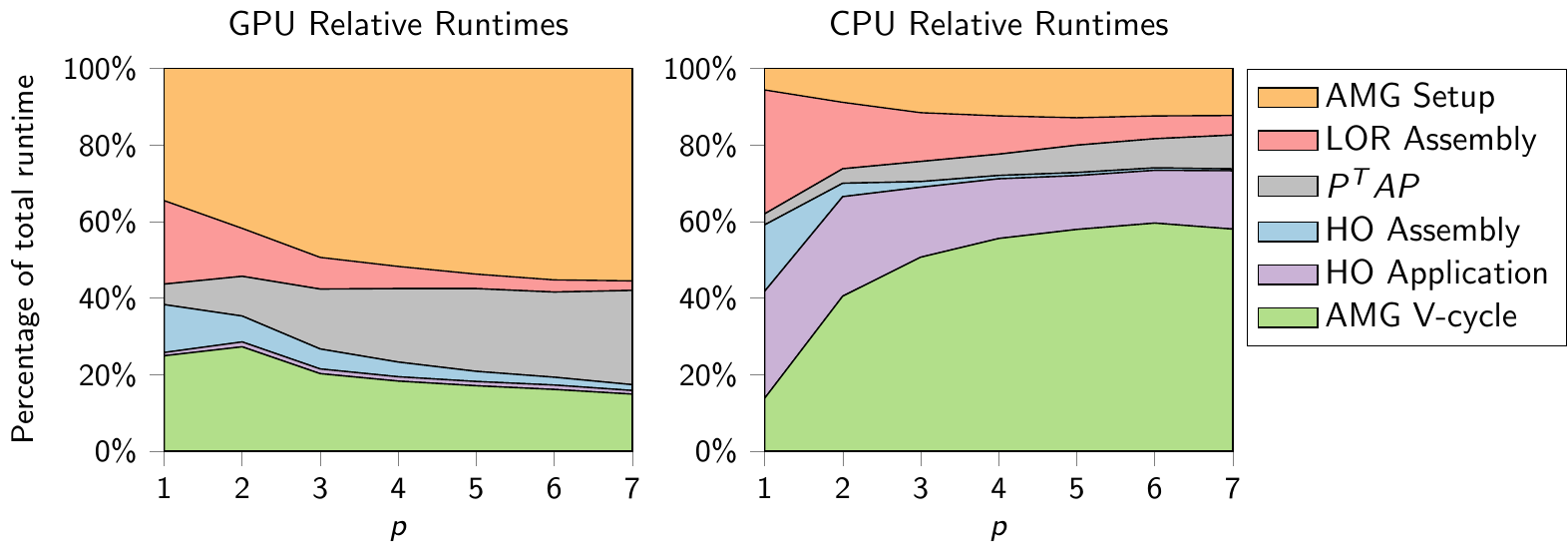}}
   \caption{
      Results for diffusion problem on mesh with nonconforming adaptive refinement.
      Relative runtime of algorithmic components, left: GPU relative runtimes (twelve V100 GPUs), right: CPU relative runtimes (one Power 9 core).
   }
   \label{fig:amr-results}
\end{figure}

We solve this problem using polynomial degrees $p=1$ through $p=7$, using three nodes of \textit{Lassen} with twelve V100 GPUs.
The relative GPU runtimes of the algorithmic components are shown in \Cref{fig:amr-results}.
For comparison, we also include the relative CPU runtimes.
As in the previous test cases, the AMG setup represents the dominant portion of the total time to solution on the GPU.
In this case, the setup is even more expensive because of the increased fill-in.
Furthermore, the triple-product $P^T A P$ (where the matrix $P$ represents both the parallel decomposition \textit{and} the nonconforming constraints) represents an increasing fraction of the total runtime as the constraint matrix becomes increasingly coupled at higher orders.
The high-order operator setup, low-order-refined matrix assembly (not including constraints enforced through the triple-product), and high-order operator application (totaled over all the CG iterations) represent 10\% or less of the time to solution for $p \geq 4$.
On the other hand, on the CPU, the relative cost of the AMG V-cycle and high-order operator evaluation are more significant.
For $p > 2$, the AMG V-cycle represents the majority of the runtime on CPU, and the high-order operator evaluation represents roughly 15\% of the CPU runtime.
In contrast to the GPU runtimes, on CPU, the AMG setup and $P^T A P$ operation each represent less than 10\% of the total runtime.

The nonconforming refinement results in nontrivial constraints that are then incorporated into the prolongation operator $\Lambda$ (cf.\ \Cref{sec:amr}), resulting in greater fill-in, and a less sparse LOR matrix.
The constraints at a nonconforming interface will generally couple all the high-order degrees of freedom lying the adjacent faces, resulting in decreased sparsity with increasing polynomial degree, see \Cref{tab:amr-results}.
However, we emphasize that despite this modest decrease in sparsity, for $p=7$ the LOR matrix $A_h$ is still roughly an order of magnitude sparser than the corresponding high-order matrix $A_p$.
The number of conjugate gradient iterations required to converge to a relative tolerance of $10^{-12}$ remains roughly constant for $p > 1$, corroborating the spectral equivalence results described in \Cref{sec:amr}.

\begin{table}
   \caption{
      Results for diffusion problem on mesh with nonconforming adaptive refinement.
      Problem size, LOR system sparsity, CG iterations required to converge to a relative residual of $10^{-12}$, and wall-clock runtime (twelve V100 GPUs on three nodes of \textit{Lassen}).
      }
   \label{tab:amr-results}

   \begin{tabular}{c|ccccc}
      \toprule
      \small $p$ & \small DOFs & \small NNZ & \small NNZ per row & \small Its. & \small GPU Runtime\,(s) \\
      \midrule
      1 & $6.0 \times 10^4$ & $1.7 \times 10^6$& 28 & 28 & 0.4 \\
      2 & $6.1 \times 10^5$ & $2.2 \times 10^7$& 36 & 43 & 0.7 \\
      3 & $2.2 \times 10^6$ & $8.8 \times 10^7$& 40 & 42 & 1.1 \\
      4 & $5.5 \times 10^6$ & $2.3 \times 10^8$& 42 & 44 & 2.0 \\
      5 & $1.1 \times 10^7$ & $5.0 \times 10^8$& 45 & 45 & 3.3 \\
      6 & $1.9 \times 10^7$ & $9.2 \times 10^8$& 48 & 46 & 5.7 \\
      7 & $3.1 \times 10^7$ & $1.6 \times 10^9$& 52 & 47 & 9.9 \\
      \bottomrule
   \end{tabular}
\end{table}

\subsection{Large-scale electromagnetic diffusion}
\label{sec:electromagnetics}

In this section, we use the GPU-accelerated solvers described above to solve a large-scale magnetic diffusion problem posed on a realistic geometry.
This problem illustrates the use of low-order-refined BoomerAMG and AMS preconditioners in $H^1$ and $\Hcurl$, and the representation of vector fields in $\Hdiv$, using the sparse discrete differential operators resulting from the interpolation--histopolation basis.
We consider a charged copper coil in air, and solve for the resulting magnetic field using the $A$--$\phi$ potential formulation of magnetic diffusion (cf.\ \cite{Rieben2006}).
The domain $\Omega$ is the box $[-\tfrac{1}{2}, \tfrac{1}{2}] \times [-\tfrac{1}{2}, \tfrac{1}{2}] \times [-\tfrac{3}{4}, \tfrac{3}{4}]$, partitioned into non-overlapping regions representing the coil and air, $\Omega = \Omega_{\text{coil}} \cup \Omega_{\text{air}}$.
The piecewise constant conductivity coefficient $\beta$ is defined as
\[
   \beta = \begin{cases}
      1 & \text{in $\Omega_{\text{coil}}$,} \\
      10^{-6} & \text{in $\Omega_{\text{air}}$.}
   \end{cases}
\]
The coil intersects the domain boundary at two terminals, $\Gin$ and $\Gout$, such that $\partial\Omega = \Gin \cup \Gout \cup \Gbox$.

The current running through the copper coil is driven by a potential difference enforced as boundary conditions at the two terminals, $\Gin$ and $\Gout$.
The electric scalar potential $\phi$ is obtained as the solution to the Poisson problem
\begin{equation}
   \label{eq:phi}
   \begin{alignedat}{3}
      \nabla \cdot \beta \nabla \phi &= 0 \quad&&\text{in $\Omega$,} \\
      \phi &= \phi_{\text{in}} \quad&&\text{on $\Gin$,} \\
      \phi &= \phi_{\text{out}} \quad&&\text{on $\Gout$,} \\
      \frac{\partial \phi}{\partial n} &= 0 \quad&& \text{on $\Gbox$.}
   \end{alignedat}
\end{equation}
For this problem, we take $\phi_{\text{in}} = 0$ and $\phi_{\text{out}} = 1$.
After obtaining the electric scalar potential, the magnetic vector potential $A$ is given as the solution to the curl-curl problem with homogeneous tangential Dirichlet conditions at all domain boundaries,
\begin{equation}
   \label{eq:A}
   \begin{alignedat}{3}
      \nabla \times \nabla \times A + \beta A &= -\beta \nabla \phi \quad&&\text{in $\Omega$,} \\
      n \times A &= 0 \quad&&\text{on $\partial\Omega$.}
   \end{alignedat}
\end{equation}
The magnetic field $B$ is given by the curl of the vector potential, $B = \nabla \times A$.

The domain is discretized using a hexahedral mesh with element boundaries fitted to the air-coil interface.
To obtain this mesh, first an unstructured tetrahedral mesh of the geometry was generated, and then each tetrahedron was subdivided into four hexahedra to obtain an all-hexahedral mesh.
While this tet-to-hex strategy often results large meshes with poorly shaped elements, it has been used successfully in challenging high-order and spectral element applications (see e.g.\ \cite{Yuan2020}).
The resulting mesh consists of 1{,}532{,}116 hexahedral elements.
We solve this problem with polynomial degree $p=4$.
The high-order $H^1$ finite element space has approximate $10^8$ degrees of freedom, and the \Nedelec space has about $2.9 \times 10^8$ degrees of freedom.

The first step of the solution procedure requires the solution of the Poisson problem \eqref{eq:phi} to compute the electric scalar potential $\phi$.
The resulting LOR system for the $H^1$ system has 27 nonzeros per row, leading to a total of $2.6\times 10^9$ nonzeros.
Relative and absolute tolerances of $10^{-8}$ were used as stopping criteria for the conjugate gradient solver, resulting in convergence after 45 iterations.
After the potential $\phi$ has been found, the right-hand side for $\eqref{eq:A}$ must be computed.
First, $\nabla \phi$ is computed by applying the discrete gradient operator, obtaining a vector field in $\Hcurl$.
This operation is performed as a sparse matrix-vector product with the discrete gradient matrix whose assembly is described in \Cref{sec:lor-assembly};
for equal problem size, the number of nonzeros in this matrix is independent of the polynomial degree.
After having computed $\nabla \phi$, the right-hand side is given by applying the $\Hcurl$ mass matrix weighted by the conductivity coefficient $\beta$.
The mass matrix is applied matrix-free using partial assembly, see \Cref{sec:pa-apply}.

\begin{figure}
   \raisebox{-0.5\height}{\includegraphics[width=2.5in]{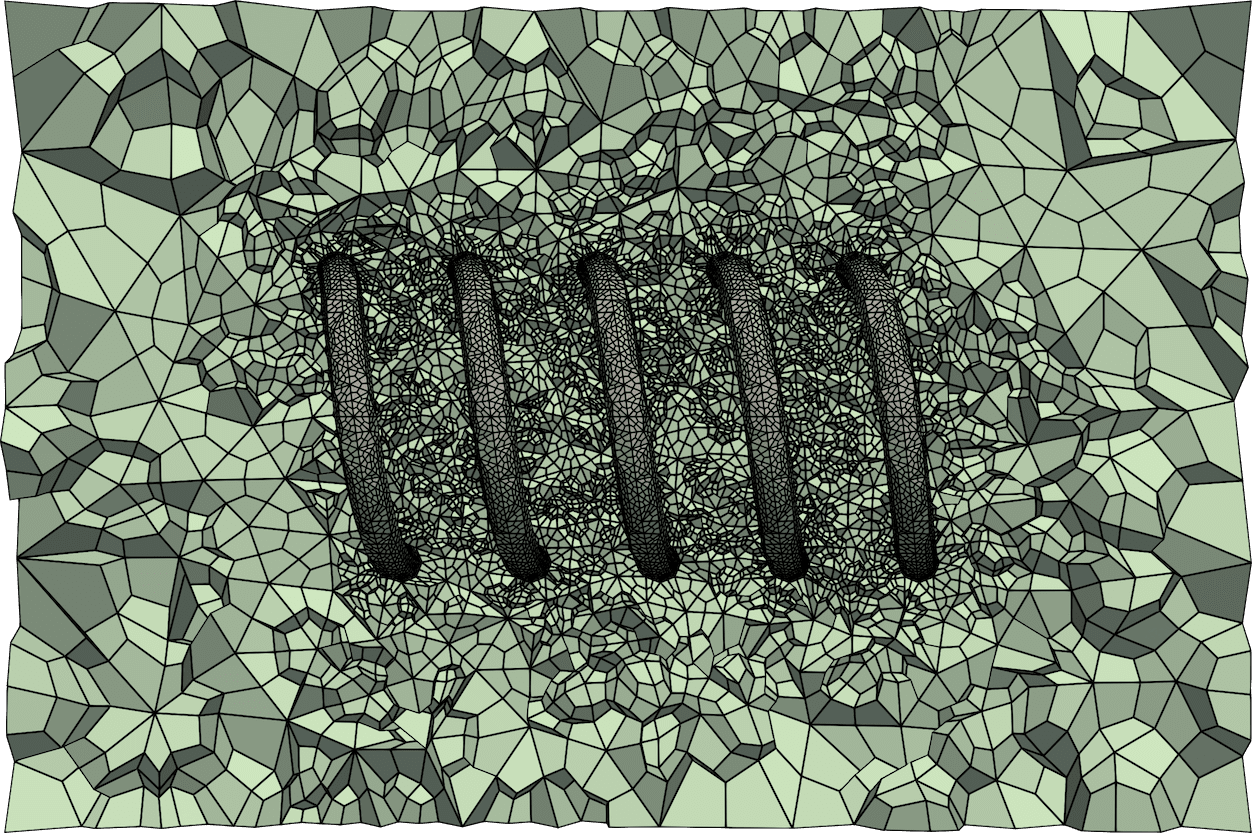}}
   \hspace{24pt}
   \raisebox{-0.5\height}{\includegraphics[width=2in]{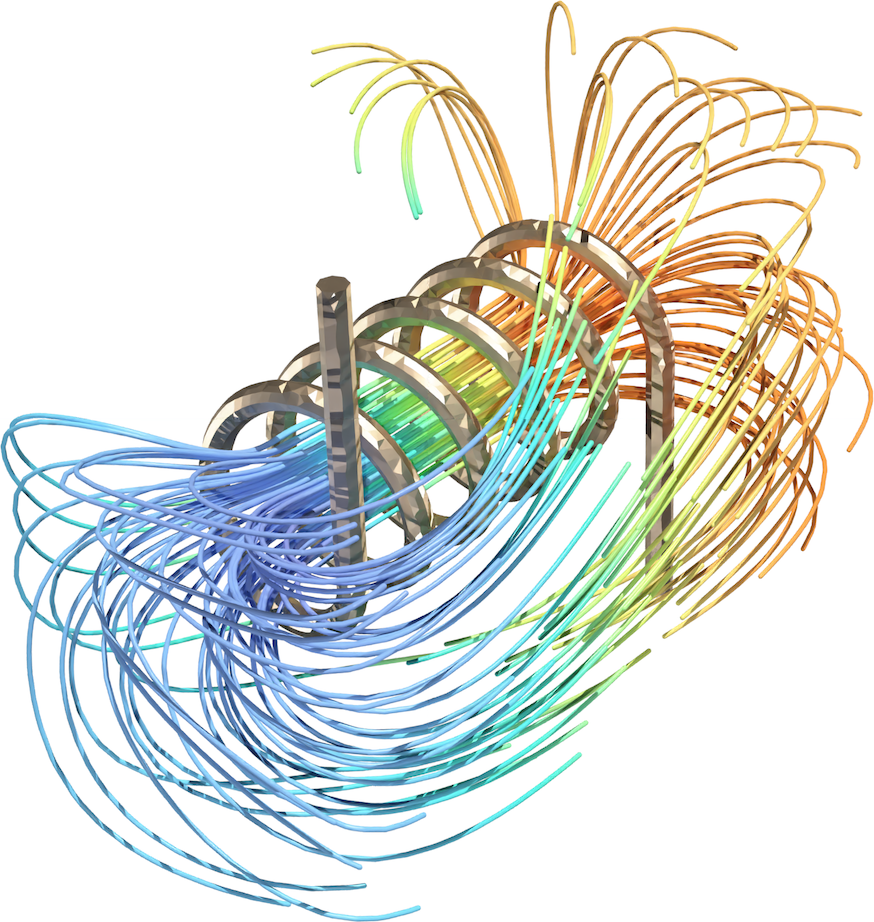}}
   \caption{
      Left: illustration of a coarser version of the mesh for the coil problem, cut with a plane passing through the origin.
      Right: magnetic field streamlines colored by electric scalar potential.
   }
   \label{fig:coil}
\end{figure}

Given the right-hand side computed above, the curl-curl problem \eqref{eq:A} is solved to obtain the magnetic vector potential $A$.
The LOR system used to construct the AMS preconditioner has approximately 33 nonzeros per row, resulting in $9.7 \times 10^9$ nonzero entries.
The same stopping criteria were used as in the $H^1$ problem.
The CG iteration converged after 22 iterations.
Once the magnetic vector potential has been computed, the magnetic field is represented as a function in $\Hdiv$ by applying the discrete curl operator.
As in the case of the discrete gradient operator, this operation is performed as a sparse matrix-vector product using the approach described in \Cref{sec:lor-assembly}.
Streamlines of the magnetic field are shown in \Cref{fig:coil}.
This problem was run on 80 nodes of \textit{Lassen} using 320 V100 GPUs.
The wall-clock runtime for the entire solution procedure was 26 seconds.

\section{Open-Source Code Availability}
\label{sec:software}

All the algorithms and implementations described in this paper are available in the open source MFEM software library under the permissive BSD license at \url{https://mfem.org} and \url{https://github.com/mfem/mfem}, see also \cite{Anderson2020,MFEM}.
The construction and use of matrix-free low-order-refined solvers is illustrated in the included \texttt{lor\_solvers} mini-application and its parallel counterpart \texttt{plor\_solvers}.
Generally, the construction of LOR solvers can be performed in one or two lines of code, and GPU acceleration and the high-performance macro-element batching strategies are automatically enabled.
Example code illustrating the construction of the low-order-refined discretization, assembling the matrix in parallel CSR format, and creating AMG preconditioners using the \textit{hypre} library is shown in \Cref{fig:code-listings}.

\begin{figure}[!h]
\begin{minipage}[t]{0.49\linewidth}
\begin{lstlisting}[language=C++]
// Set up bilinear form 'a' and the list of essential
// degrees of freedom 'ess\_dofs'
// [$\cdots$]

// Form the low-order-refined discretization, and assemble
// the LOR matrix in parallel CSR format
LORDiscretization lor(a, ess_dofs);
HypreParMatrix &lor_matrix =
   lor.GetAssembledMatrix();
\end{lstlisting}
\end{minipage}\hfill
\begin{minipage}[t]{0.49\linewidth}
\begin{lstlisting}[language=C++]
// For any SolverType (AMG, direct solver, etc.), form the
// corresponding LOR preconditioner
LORSolver<SolverType> lor_solver(a, ess_dofs);

// For example:
// if 'a' is H1 diffusion...
LORSolver<HypreBoomerAMG> lor_amg(a, ess_dofs);
// if 'a' is ND curl-curl...
LORSolver<HypreAMS> lor_ams(a, ess_dofs);
// if 'a' is RT div-div...
LORSolver<HypreADS> lor_ads(a, ess_dofs);
\end{lstlisting}
\end{minipage}
\caption{
Illustration of MFEM's API for the construction and use of the low-order-refined solvers.
Left: forming the low-order-refined discretization and assembling the matrix in parallel CSR format.
Right: creating low-order-refined \textit{hypre} AMG, AMS, and ADS preconditioners given the high-order bilinear form.
}
\label{fig:code-listings}
\end{figure}

\section{Conclusions}
\label{sec:conclusions}

In this paper we have described the algorithms and implementations for the GPU-accelerated solution of high-order finite element problems using low-order-refined preconditioning.
This approach combines matrix-free operator evaluation using the partial assembly approach with algebraic multigrid preconditioners constructed using assembled low-order-refined matrices.
The end-to-end GPU acceleration of LOR preconditioning allows for the efficient and scalable solution to high-order finite element problems on GPU-based supercomputing architectures, avoiding the memory-intensive and costly assembly of the high-order system matrices.
New LOR matrix assembly algorithms based on macro-element batching were introduced, achieving significant speedup when compared with fully unstructured low-order matrix assembly.
A detailed study of kernel throughput for the algorithmic components of the solution procedure was presented.
The performance of preconditioners on problems with nonconforming adaptive mesh refinement is considered, showing fast problem setup and convergence even at high orders.
The scalability of the solvers was demonstrated on problems with over 1 billion degrees of freedom on 1024 GPUs.
Finally, we demonstrated the capability of these solvers on a challenging large-scale electromagnetic diffusion problem with complex geometry using 320 GPUs.

\section{Acknowledgments}

The authors thank V.\ Dobrev for the unfailingly insightful comments and suggestions, and M.\ Stowell for help with the problem formulation in \Cref{sec:electromagnetics}.
This work was performed under the auspices of the U.S.\ Department of Energy by Lawrence Livermore National Laboratory under Contract DE-AC52-07NA27344 (LLNL-JRNL-841564).
W.\ Pazner and Tz.\ Kolev were partially supported by the LLNL-LDRD Program under Project No.~20-ERD-002.
This research was supported by the Exascale Computing Project (17-SC-20-SC), a collaborative effort of two U.S. Department of Energy organizations (Office of Science and the National Nuclear Security Administration) responsible for the planning and preparation of a capable exascale ecosystem, including software, applications, hardware, advanced system engineering and early testbed platforms, in support of the nation's exascale computing imperative.

\printbibliography

\end{document}